\documentclass[numreferences]{kluwer}                     
\usepackage{amsfonts}
\newtheorem{lemma}{Lemma}
\newtheorem{satz}{Proposition}
\newtheorem{theorem}{Theorem}
\newtheorem{corolla}{Corollary}
\newdisplay{problem}{Problem}
\newdisplay{defi}{Definition}
\newdisplay{conj}{Conjecture}
\newdisplay{exam}{Example}
\newdisplay{cexam}{Counterexample}
\newdisplay{notation}{Notations}
\newdisplay{remark}{Remark}

\begin{document}
\begin{article}
\begin{opening}
\title{On Bures-Distance and $^*$-Algebraic Transition Probability 
between Inner Derived Positive Linear Forms over\\  
${\mathsf W}^*$-Algebras}
\author{Peter M.\surname{\,Alberti}\email{Peter.Alberti@itp.uni-leipzig.de}
\thanks{Partially supported by `Deutsche Forschungsgemeinschaft'.}\\}
\author{Armin \surname{Uhlmann}\email{Armin.Uhlmann@itp.uni-leipzig.de}}
\institute{Institute of Theoretical Physics\\
University of Leipzig\\
Augustusplatz 10, D-04109 Leipzig, Germany}
\runningtitle{Bures-Distance between Inner Derived Positive Linear Forms}
\runningauthor{P.\,Alberti and A.\,Uhlmann}
\begin{ao}
Institute of Theoretical Physics, University of Leipzig, 
Augustusplatz 10, D-04109 Leipzig, Germany
\end{ao}
\begin{abstract}
On a ${\mathsf W}^*$-algebra $M$, for given two positive linear forms 
$\nu,\varrho\in M_+^*$ and algebra elements $a,b\in M$ a variational expression  
for the Bures-distance $d_{\mathrm{B}}(\nu^a,\varrho^b)$ between 
the inner derived positive linear forms  
$\nu^a=\nu(a^*\cdot\,a)$ and $\varrho^b=\varrho(b^*\cdot\,b)$ is obtained. 
Along with the 
proof of the formula also some earlier result of 
S.\,Gudder on non-commutative probability will be slighly extended. 
Also, the given expression of the Bures-distance nicely    
relates to some system of seminorms proposed by D.\,Buchholz and which 
occured along with the problem of estimating the 
so-called `weak intertwiners' in algebraic quantum field theory. In the 
last part some optimization problem will be considered.      
\end{abstract}
\classification{MSC codes}{46L89, 46L10, 58B10, 58B20}
\date{}
\end{opening}
\section{Introduction}\label{intro}
\subsection{Basic settings on Bures-distance}\label{intro.1}
Throughout the paper the  
Bures-distance function $d_{\mathrm{B}}$ \cite{Bure:69} and related metric concepts on the 
positive cone $M_+^*$ of the bounded linear forms $M^*$ over a 
${\mathsf W}^*$-algebra $M$ will be considered. 
Start with defining the Bures-distance $d_{\mathrm{B}}(M|\nu,\varrho)$
between $\nu,\varrho\in M_+^*$.
\begin{defi}\label{budi} 
$\ \ d_{\mathrm{B}}(M|\nu,\varrho)=\inf_{\{\pi,{\mathcal K}\},\varphi\in 
{\mathcal S}_{\pi,M}(\nu),\psi\in {\mathcal S}_{\pi,M}(\varrho)} 
\|\psi-\varphi\|\,.$
\end{defi}
Instead of $d_{\mathrm{B}}(M|\nu,\varrho)$ often also    
$d_{\mathrm{B}}(\nu,\varrho)$ will be used.   
For unital $^*$-representation $\{\pi,{\mathcal K}\}$ of $M$ on a Hilbert space $\{{\mathcal K},<\cdot,\cdot>\}$ 
and for $\mu\in M_+^*$ we let ${\mathcal S}_{\pi,M}(\mu)=\{\xi\in {\mathcal K}:\mu(\cdot)=
\langle\pi(\cdot)\xi,\xi\rangle\}$. Then, the infimum within the defining formula for 
$d_{\mathrm{B}}(\nu,\varrho)$ extends  
over all those $\pi$ relative to which ${\mathcal S}_{\pi,M}(\nu)\not=\emptyset$ and 
${\mathcal S}_{\pi,M}(\varrho)\not=\emptyset$ simultaneously hold, and within each 
such representation the vectors $\varphi$ and $\psi$ may be varied 
through all of ${\mathcal S}_{\pi,M}(\nu)$ and ${\mathcal S}_{\pi,M}(\varrho)$, respectively.    
The scalar product ${\mathcal K}\times {\mathcal K}
\ni\{\chi,\eta\}\,\longmapsto\,\langle\chi,\eta\rangle\in 
{\mathbb{C}}$ on the representation Hilbert space by convention is supposed 
to be linear with respect to the first argument $\chi$, and antilinear
in the second argument $\eta$, and maps into the complex 
field ${\mathbb{C}}$. Let ${\mathbb{C}}\ni z\mapsto \bar{z}$ be the complex conjugation, 
and be $\Re z$ and $|z|$ the real part and absolute value of $z$, respectively. Greek letters  
and their labelled derivates (except for $\pi$, which is reserved for representations only) 
will be used to label elements of the complex Hilbert spaces on which the
concrete ${\mathsf C}^*$-algebras $\pi(M)$ are 
supposed to act. The norm of $\chi\in
{\mathcal K}$ is given by $\|\chi\|=\sqrt{\langle\chi,\chi\rangle}$.  
Relating operator and ${\mathsf C}^*$-algebra theory, refer the reader to standard 
monographs, 
e.g.~\cite{Dixm:64,Saka:71,KaRi:83}.

For simplicity, for the ${\mathsf C}^*$-norm of an element 
$x\in M$ as well as for the operator norm of an concrete bounded linear 
operator $x\in {\mathsf B}({\mathcal K})$ the same notation 
$\|x\|$ will be used. In both these cases, the involution ($^*$-operation) 
respectively the taking of the hermitian conjugate of an element $x$ is 
indicated by the transition $x\,\longmapsto\,x^*$. The notions of hermiticity and positivity 
for elements are defined as usual in ${\mathsf C}^*$-algebra theory, and 
$M_{\mathrm{h}}$ and $M_+$ are the hermitian and positive elements of $M$, 
respectively. With view to the above, to make these settings more 
unambiguous, agree that greek letters will {\em not} be used as
symbols for linear operators over ${\mathcal K}$ or elements of $M$. The null  
and the unit element/operator in $M$ and ${\mathsf B}({\mathcal 
K})$ will be denoted by ${\mathbf 0}$ and ${\mathbf 1}$. 

For notational purposes mainly, in short recall some   
fundamentals relating (bounded) linear forms which subsequently 
might be of concern in context of {\em Definition \ref{budi}}. Remind that the 
topological dual space $M^*$ of $M$ is the
set of all those linear functionals (linear forms) which are continuous with respect to the 
operator norm topology. Equipped with the dual norm
$\|\cdot\|_1$, which is
given by $\|f\|_1=\sup\{|f(x)|\,:\,x\in M,\,\|x\|\leq 1\}$ 
and which is referred to as the functional norm, $M^*$ is a Banach space. 
For each given $f\in M^*$, the hermitian
conjugate functional $f^*\in {M}^*$ is defined by $f^*(x)=\overline{f(x^*)}$, for
each $x\in M$. Remind that $f\in M^*$ is hermitian 
if $f=f^*$ holds, and $f$ is termed positive if $f(x)\geq 0$ holds, 
for each $x\in M_+$. Also remind that a bounded linear form over $M$ is positive 
if, and only if, $\|f\|_1=f({\mathbf 1})$ is fulfilled. For positive linear 
forms one has the following fundamental estimate (Cauchy-Schwarz inequality)
\begin{subequation}
\begin{equation}\label{cauchy}
\forall\,g\in M_+^*\,:\ |g(y^*x)|^2\leq g(y^*y) \, 
g(x^*x),\,\forall\,x,y\in M\,, 
\end{equation} 
which accordingly also holds on ${\mathsf C}^*$-algebras. From this it is 
easily inferred that for each $g\in M_+^*\backslash \{{\mathbf 0}\}$ the subset $I_g\subset M$  
defined by 
\begin{equation}\label{ide}
I_g=\{ x\in M:\ g(x^*x)=0\} 
\end{equation}
\end{subequation} 
is a (proper) {\em left ideal}\, in $M$. Provided this ideal is trivial, $I_g=\{{\mathbf 0}\}$, 
the positive linear form 
$g\in M_+^*$ is called {\em faithful} (positive linear form). 

The most important consequence of positivity and (\ref{cauchy}) is that, for 
each $g\in M_+^*$, there exists a cyclic $^*$-representation 
$\pi_g$ of $M$ on some Hilbert space ${\mathcal K}_g$, with cyclic vector 
${\mathit \Omega}\in {\mathcal K}_g$, and obeying $g(x)=\langle \pi_g(x){\mathit \Omega},
{\mathit \Omega}\rangle$, for all $x\in M$. This fact usually is referred to as   
the Gelfand-Neumark-Segal theorem (GNS). Such a representation (which is unique up to unitary 
isomorphisms) will be referred to as 
with $g$ associated cyclic representation, or GNS-representation of $g$, 
respectively. Note that considering such construction in the special case 
with $g=\nu+\varrho$ will provide a unital $^*$-representation $\pi=\pi_g$ 
such that ${\mathcal S}_{\pi,M}(\nu)\not=\emptyset$ and ${\mathcal S}_{\pi,M}(\varrho)\not=
\emptyset$ hold (we omit the details, all of which are standard). It is 
exactly this fact which makes that the expression in {\em Definition 
\ref{budi}} makes sense even in the ${\mathsf C}^*$-algebraic case.

Apart from the functional norm topology, mention also 
the $w^*$-topology on $M^*$, which is the weakest locally convex topology 
generated by the seminorms $\rho_x$, $x\in M$, with $\rho_x(f)=|f(x)|$, 
for each $f\in M^*$. Remind that according to a basic result of Banach space theory 
(Alaoglu-Banach theorem) each closed, bounded subset of the dual Banach space $M^*$ 
has to be $w^*$-compact. 
 
Along with {\em Definition \ref{budi}} an auxiliary 
metric structure arises which can be compared to the metric 
structure given by the `natural' distance  
$d_1(\nu,\varrho)=\|\nu-\varrho\|_1$ on $M_+^*$. The relevant  
basic facts will be stated here without proof and read as follows.  
\begin{satz}\label{budi0}
Let $d_{\mathrm{B}}\,:\ M_+^*\times
M_+^*\ni\{\nu,\varrho\}\,\longmapsto\,d_{\mathrm{B}}(M|\nu,\varrho)\in
{\mathbb{R}}_+$ be given in accordance with Definition \ref{budi}. 
Then the following hold\textup{:} 
\begin{enumerate}[0000]
\item\label{budi0.1}
$d_{\mathrm{B}}$ is a distance function on the points of $M_+^*$;
\item\label{budi0.2}
$d_{\mathrm{B}}$ is topologically equivalent with 
$d_1$ on bounded subsets of $M_+^*$. 
\end{enumerate}
Especially, for $\{\nu,\varrho\}\in M_+^*\times M_+^*\backslash \{0,0\}$ one has 
\begin{equation}\label{7.6b}
c(\nu,\varrho)^{-1}\,d_1(\nu,\varrho) \leq  d_{\mathrm{B}}(M|\nu,\varrho) \leq 
\sqrt{d_1(\nu,\varrho)}\,,
\end{equation}
with $c(\nu,\varrho)=\sqrt{\|\nu\|_1}+\sqrt{\|\varrho\|_1}$.
\end{satz}
Remark that item (\ref{budi0.1}) 
and `one half' of the estimate (\ref{7.6b}), from which  
(\ref{budi0.2}) obviously can be followed, were  
anticipated and proved by D.\,Bures in \cite{Bure:69}, whereas the other 
half of (\ref{7.6b}) can be seen by arguments given 
by H.\,Araki in \cite{Arak:71,Arak:72} e.g.; omit any details on this 
matter but remark that D.\,Bures refers to the {\em state space} of $M$,  
${\mathcal S}(M)=\{f\in M_+^*:f({\mathbf 1})=1\}$. This simplifies matters 
insofar that in restriction to ${\mathcal S}(M)$   
then $d_{\mathrm{B}}$ gets unconditionally 
topologically equivalent with 
$d_1$.  
 
\subsection{Prerequisites, useful estimates and examples}\label{intro.2}
In conjunction with the Bures-distance $d_{\mathrm B}$ one has the functor $P$ of the  
($^*$-algebraic) {\em transition probability} \cite{Uhlm:76}. 
For given ${\mathsf W}^*$-algebra $M$ and positive 
linear forms $\nu,\varrho\in M_+^*$ the definition reads as follows\,:
\begin{defi}\label{genprob.2}
\ \ $P_M(\nu,\varrho)=
\sup_{\{\pi,{\mathcal K}\},\varphi\in 
{\mathcal S}_{\pi,M}(\nu),\psi\in {\mathcal S}_{\pi,M}(\varrho)}
|\langle \psi,\varphi\rangle|^2\,.$
\end{defi}
Thereby, the range of variables over which the supremum has to 
be extended is the same as in {\em Definition \ref{budi}}. 
With the help of $P_M$ one then gets a (uniquely solvable) expression 
for the Bures-distance\,: 
\begin{equation}\label{pcont.1}
d_{\mathrm{B}}(M|\nu,\varrho)^2=\biggl\{\|\nu\|_1-
\sqrt{P_M(\nu,\varrho)}\biggr\} +
\biggl\{\|\varrho\|_1-\sqrt{P_M(\nu,\varrho)}\biggr\}\,. 
\end{equation}
Remark that $P$ is of importance on its own rights (and independent from the just mentioned 
appearance within (\ref{pcont.1})) since it can be well-adapted to several  
applications in (algebraic) quantum physics, non-commutative probability and 
estimation theory. The latter also was the heuristic intention behind 
the introduction of this functor in \cite{Uhlm:76}. 
For a particular range of applications see e.g.~\cite{AlHe:89,Uhlm:95}.

Many properties of $P$ are known. In the following only some very 
few of these properties will be referred to explicitely. For instance, essentially by means 
of the Cauchy-Schwarz inequality from the definition of $P$ the 
following fundamental estimates can be obtained\,:
\begin{equation}\label{au76}
|f({\mathbf 1})|^2\leq P_M(\nu,\varrho)\leq \nu(a)\,\varrho(a^{-1})\,,
\end{equation}
where $f$ can be any linear form of the set 
\begin{equation}\label{GCS}
\Gamma_M(\nu,\varrho)=\biggl\{ f\in M^*: \bigl|f(y^*x)\bigr|^2\leq 
\nu(y^*y)\varrho(x^*x),\,\forall \,x,y\in M\biggr\}  
\end{equation} 
and \,$a$\, can be any invertible, positive element $a\in M_+$. Note that 
$\Gamma_M(\nu,\varrho)$ obviously is $w^*$-closed and bounded 
($\sqrt{{\|\nu\|_1\|\varrho\|_1}}$ is a common upper bound), 
and thus is a $w^*$-compact subset of $M^*$. 

For the estimate from above see eq.\,(16) in \cite{Uhlm:76}. Relating the estimate from below, 
suppose a unital $^*$-representation 
$\{\pi,{\mathcal K}\}$ of $M$ on ${\mathcal K}$  
with ${\mathcal S}_{\pi,M}(\nu)\not=\emptyset$ and  
${\mathcal S}_{\pi,M}(\varrho)\not
=\emptyset$ to be given. By standard facts one then infers that for 
given $\varphi\in {\mathcal S}_{\pi,M}(\nu)$,  
$\psi\in {\mathcal S}_{\pi,M}(\nu)$ 
\begin{equation}\label{schief0}
\Gamma_M(\nu,\varrho)=\biggl\{\langle\pi(\cdot)k\psi,\varphi\rangle\,:\,
k\in (\pi(M)^{\,\prime})_1\biggr\}
\end{equation}
has to be fulfilled. In this formula $(\pi(M)^{\,\prime})_1$ is the unit ball within the 
commutant $vN$-algebra $\pi(M)^{\,\prime}$. From this and 
{\em Definition \ref{genprob.2}} with the help of the Theorem of 
B.\,Russo and H.\,Dye \cite{DyRu:66} then also the validity of 
the estimate from below in (\ref{au76}) follows, see eq.\,(3) in \cite{Albe:83}. 

Apply (\ref{au76}) to the special case of two vector states, which is heuristically 
important in a quantum physical context of two wave functions\,:
\begin{exam}\label{pure}
Let $M={\mathsf B}({\mathcal H})$ be the algebra of bounded linear operators 
on a Hilbert space ${\mathcal H}$. Let $\mu_\psi=\langle(\cdot)\psi,\psi\rangle$ be the vector form 
generated by $\psi\in {\mathcal H}$ on $M$, and be $p_\varphi$ the orthoprojection 
onto the span of $\varphi\in {\mathcal H}$. 
Then, considering $f=\langle(\cdot)\psi,\varphi\rangle\in 
\Gamma_M(\mu_\varphi,\mu_\psi)$ and 
$a=p_\varphi+\varepsilon^{-1} p_\varphi^\perp$, for $\varepsilon\in 
{\mathbb R}_+\backslash \{0\}$, and inserting this into 
(\ref{au76}) provides 
$|\langle\psi,\varphi\rangle|^2\leq P_M(\mu_\varphi,\mu_\psi)\leq 
|\langle\psi,\varphi\rangle|^2 + \varepsilon
\|\varphi\|^2\mu_\psi(p_\varphi^\perp)$. For $\varepsilon\to 0$ from this 
$$P_M(\mu_\varphi,\mu_\psi)=|\langle\psi,\varphi\rangle|^2$$ 
follows, in any case of two vectors $\psi,\varphi\in {\mathcal H}$.   
\end{exam}
Also, constellations among the positive 
linear forms $\nu,\varrho\in M_+^*$ are known such that for some $a\geq {\mathbf 0}$ the upper estimate 
within (\ref{au76}) turns into 
an equality. This then provides an expression for $P_M(\nu,\varrho)$.

To explain such stuff, fix some notation first. In all that follows for  
$x\in M$ and $\mu\in M_+^*$ a positive linear form $\mu^x$ 
will be defined by $\mu^x(y)=\mu(x^*y x)$, for each $y\in M$. If this 
situation occurs the positive linear form $\mu^x$ will be referred to as 
an {\em inner derived} (from $\mu$) positive linear form. 
The main result of \cite{Uhlm:76} refers to this and 
reads as follows\,:
\begin{theorem}\label{uhlrem.2}
$\forall\,\mu\in M_+^*,\,a,b\in M,\,a^*b\geq {\mathbf 0}\,:\,  
\sqrt{P_M(\mu^a,\mu^b)}=\mu(a^*b).$
\end{theorem}
For instance, in chosing $a\geq {\mathbf 0},\,b={\mathbf 1}$ 
the premises of the previous result are fulfilled in a trivial manner and 
one thus arrives at the formula 
\begin{equation}\label{au76a} 
\forall\,\mu\in M_+^*,\,a\in M_+\,:\ \ P_M(\mu^a,\mu)=\mu(a)^2\,.
\end{equation}
Remark that {\em Example \ref{pure}} in the case of non-orthogonal 
vectors can be seen as a special case of (\ref{au76a}) as well.  
It is interesting that the seemingly very special situation with  
the premises of (\ref{au76a}) addresses itself to a wide range of 
characteristic applications. One of these reads as follows\,: 
\begin{exam}\label{rad}
By the Radon-Nikodym theorem of S.\,Sakai \cite{Saka:65} we are always 
in such a situation if amongst two {\em normal} positive 
linear forms $\nu,\varrho\in M_+^*$ 
a relation of domination $\varrho\leq \lambda\/\nu$, with 
$\lambda\in {\mathbb R}_+\backslash \{0\}$, takes place, in which situation also the notation 
$\varrho\ll \nu$ will be used. That is, for $\varrho\ll \nu$ there is $a\in M_+$ 
with $\varrho=\nu^a$. In view of the 
above in such situation then especially $P_M(\varrho,\nu)=\nu(a)^2$ follows. 
It is known that $a$ gets unique if    
$s(a)\leq s(\nu)$ is required to hold, with the supports of the operator $a$ and 
normal positive linear form $\nu$, respectively. To this unique 
\,$a$\, one usually refers as the Sakai's Radon-Nikodym operator of $\varrho$ relative to $\nu$,  
and then also the notation $a=\sqrt{d\/\varrho/d\/\nu}$ will be used.   
\end{exam}  

Finally, it is interesting that in any case with the help of the bounds 
appearing along with (\ref{au76}) the value of 
$P_M(\nu,\varrho)$ can be 
approximated to an arbitrary degree of precision from both sides. 
This and some other relevant informations 
are the content of the following result.     
\begin{theorem}\label{genprob}
Let $M$ be a ${\mathsf W}^*$-algebra, and be $\nu,\varrho\in M_+^*$.  
Then, the following facts hold\textup{:}
\begin{enumerate}[0000]
\item\label{genprob.1}
$\sqrt{P_M(\nu,\varrho)}=\inf_{x>{\mathbf 0}}\sqrt{\nu(x)\varrho(x^{-1})}\,;$
\item\label{uhl}
$\sqrt{P_M(\nu,\varrho)}=\sup_{f\in \Gamma_M(\nu,\varrho)} |f({\mathbf 1})|\,.$
\end{enumerate}
The infimum in \textup{(\ref{genprob.1})} extends over all 
positive invertible elements of $M$.  
Moreover, if $\{\pi,{\mathcal K}\}$ is any unital $^*$-representation of 
$M$ over some Hilbert space 
${\mathcal K}$ such that ${\mathcal S}_{\pi,M}(\nu)\not=\emptyset$ and     
${\mathcal S}_{\pi,M}(\varrho)\not=\emptyset$ are fulfilled, 
then the following is fulfilled\textup{:}
\begin{enumerate}[0000]
\setcounter{enumi}{2}
\item\label{genprob.3}
$\sqrt{P_M(\nu,\varrho)}=
\sup_{\psi\in {\mathcal S}_{\pi,M}(\varrho)}|\langle 
\psi,\varphi\rangle|\,,\,\forall\,\varphi\in {\mathcal S}_{\pi,M}(\nu)\,.$  
\end{enumerate}
Also, the supremum in \textup{(\ref{uhl})} 
is a maximum and is attained at some $f\in \Gamma_M(\nu,\varrho)$, and   
some maximizing 
$f$ can be chosen as $f=\langle\pi(\cdot)\psi_0,\varphi_0\rangle$, for some  
$\psi_0\in {\mathcal S}_{\pi,M}(\varrho),\,\varphi_0
\in {\mathcal S}_{\pi,M}(\nu)$. 
\end{theorem}
For proofs of (\ref{genprob.1})--(\ref{genprob.3}) see 
{\sc Corollary 1}, {\sc Corollary 3} and {\sc Theorem 3} in 
\cite{Albe:83}, for the additional informations on the attainability of the 
supremum in (\ref{uhl}), see \cite{Arak:72} and \cite{Albe:92.1}. 
The previous result remains valid even if $M$ is supposed to be a unital 
${\mathsf C}^*$-algebra. 
\begin{remark}\label{uhlrem.0}
The question arises whether the functor $P$ in a reasonable manner (i.e.~such that a relation of 
type (\ref{pcont.1}) with a metric 
distance $d_{\mathrm B}$ remained true on its domain of definition) could be 
extended to some yet more general category of $^*$-algebras (including 
some unbounded operator algebras showing up in relativistic 
quantum field theory e.g.), see \cite{Uhlm:74,Uhlm:85}.  Besides the just mentioned 
${\mathsf C}^*$-algebraic cases the answer seems to be in the negative. 
\end{remark}
\newpage
\subsection{The main result}\label{intro.3}
Under the premises of {\sc Theorem \ref{uhlrem.2}}, let us suppose now 
that some unital $^*$-representation $\{\pi,{\mathcal K}\}$ has been chosen in 
accordance with ${\mathcal S}_{\pi,M}(\mu)\not=\emptyset$. Then, for ${\mathit \Omega}\in
{\mathcal S}_{\pi,M}(\mu)$ one has $\pi(a){\mathit \Omega}\in {\mathcal S}_{\pi,M}(\mu^a)$ and 
$\pi(b){\mathit \Omega}\in {\mathcal S}_{\pi,M}(\mu^b)$. 
Hence, in making use of  
(\ref{schief0}) in the special case of $\Gamma_M(\mu,\mu)$, with $\varphi=\psi=
{\mathit \Omega}$, and in the special case of $\Gamma_M(\mu^a,\mu^b)$ 
with $\psi=\pi(b){\mathit \Omega}$ and $\varphi=\pi(a){\mathit \Omega}$, and 
respecting positivity of \,$a^*b$, one easily infers that
$$
\mu(a^*b)=\bigl\|\pi(\sqrt{a^*b}){\mathit \Omega}\bigr\|^2=\sup_{g\in \Gamma_M(\mu,\mu)} |g(a^*b)|=
\sup_{f\in \Gamma_M(\mu^a,\mu^b)} |f({\mathbf 1})|
$$
has to be fulfilled. The formula of {\sc Theorem \ref{uhlrem.2}} and 
{\sc Theorem \ref{genprob}}\,(\ref{uhl}) together with the previous then show that the 
following is valid.
\begin{corolla}\label{premain}
$$\forall\,\mu\in M_+^*,\ a,b\in M,\ a^*b\geq {\mathbf 0}\ :\ \ 
\sqrt{P_M(\mu^a,\mu^b)}=\sup_{f\in \Gamma_M(\mu,\mu)} |f(a^*b)|\,.$$
\end{corolla}

The first goal of the paper will be to extend the assertion of  
{\sc Corollary \ref{premain}} as to hold true under much weaker 
premises.   
More precisely, instead of considering 
two positive linear forms $\nu,\,\varrho$ which both are inner derived 
positive linear forms $\nu=\mu^a$ and $\varrho=\mu^b$ from one and 
the same positive linear form $\mu$ via operators $a,b\in M$ which obey 
the positivity assumption $a^*b\geq {\mathbf 0}$, subsequently 
two arbitrarily chosen inner derived positive linear forms are permitted 
to be considered, without any further restriction. Based on this, under the same premises on the 
positive linear forms a variational expression for the Bures-distance function will 
be derived. 
\begin{theorem}\label{buch00}
Let $M$ be a ${\mathsf W}^*$-algebra, and be $\nu,\varrho\in M_+^*$, 
and $a,b\in M$. Then, the following facts hold true\textup{:}
\begin{enumerate}[0000] 
\item\label{buch.20}
$\sqrt{P_M\bigl(\nu^a,\varrho^b\bigr)}=
\sup_{f\in \Gamma_M(\nu,\varrho)} |f(a^*b)|\,;$
\item\label{uhlbuch.1}
$d_{\mathrm{B}}(M|\nu^a,\varrho^b)^2=\sup_{a^*b=y^*x} \bigl\{ 
\nu(a^*a-y^*y)+\varrho(b^*b-x^*x)\bigr\}\,.
$
\end{enumerate}
\end{theorem}

Obviously, (\ref{buch.20}) is the announced extension of the assertion of  
{\sc Corollary \ref{premain}}, whereas by (\ref{uhlbuch.1}), which will 
be shown to be a consequence of (\ref{buch.20}), the mentioned 
variational expression for the distance $d_{\mathrm{B}}$ between two inner derived from a 
given pair $\{\nu,\varrho\}$ positive linear forms is given. 

Foremost,  
such expression as given in (\ref{uhlbuch.1}) can be useful 
since it allows for estimating the 
behavior of the Bures-distance at 
$\{\nu,\varrho\}$ if this pair is undergoing an inner perturbation towards 
another pair $\{\nu^a,\varrho^b\}$ of positive linear forms. As it comes out, 
the geometry of submanifolds of 
mutually coordinated (via inner operations) positive linear forms of 
${\mathsf W}^*$-algebras to a great deal can be based on this formula. 
We will not elaborate on this in this paper, but instead within {\sc Section \ref{minimal}} 
we will be concerned with one particular aspect of this geometry more in detail.
            
In the course of the derivation of the main result several 
further characterizations of $P$ 
(and thus of $d_{\mathrm B}$ as well) will be obtained. 

\setcounter{equation}{0}
\section{Results and proofs}\label{beweise}
\subsection{Further characterizations of transition probability}\label{beweise.1}
In all what follows\endnotemark $M$ is a ${\mathsf W}^*$-algebra, and $\nu,\varrho\in M_+^*$ are 
fixed but can be arbitrarily chosen positive linear forms. Start with some consequences 
from {\sc Theorem \ref{genprob}}. Relating notations, when occuring 
in conjunction with $\inf$ or $\sup$,  in each case of occurrence  
the variables ${x>{\mathbf 0}}$, ${\{x\}}$, ${\{e\}}$ and ${\{y,x\}}$ are thought to extend  
over all positive invertible elements $x$, all finite 
decompositions $\{x\}=\{x_1,\ldots,x_n\}$ of the unity into positive elements, 
all finite decompositions 
$\{e\}=\{e_1,\ldots,e_n\}$ of the 
unity into orthoprojections, and all finite double systems 
$\{y,x\}=\{y_1,x_1,\ldots,y_n,x_n\}$ of elements obeying $\sum_j y_j^*x_j={\mathbf 1}$, 
respectively, within $M$, where $n$ can range through the naturals, $n\in {\mathbb N}$.  
\begin{corolla}\label{genproba}
The following properties hold\textup{:} 
\begin{enumerate}[0000]
\item\label{genprob.1a}
$\sqrt{P_M(\nu,\varrho)}=\inf_{\{x\}} \sum_j 
\sqrt{\nu(x_j)\varrho(x_j)}\,;$
\item\label{genprob.1b}
$\sqrt{P_M(\nu,\varrho)}=\inf_{\{e\}} \sum_j 
\sqrt{\nu(e_j)\varrho(e_j)}\,;$
\item\label{genprob.1c}
$\sqrt{P_M(\nu,\varrho)}=\inf_{\{y,x\}} \frac{1}{2}\,\sum_j 
\bigl\{\nu(y_j^*y_j)+\varrho(x_j^*x_j)\bigr\}\,;$ 
\item\label{genprob.1cc}
$\sqrt{P_M(\nu,\varrho)}=\inf_{\{{\mathbf 1}=y^*x\}} \frac{1}{2}\, 
\bigl\{\nu(y^*y)+\varrho(x^*x)\bigr\}\,;$
\item\label{genprob.1u}
$\sqrt{P_M(\nu,\varrho)}=\inf_{x>{\mathbf 0}}
\frac{1}{2}\,\bigl\{\nu(x)+\varrho(x^{-1})\bigr\}\,.$
\end{enumerate}
\end{corolla}
\bigskip   
\begin{pf}
Note that according to (\ref{GCS}) for each
$f\in \Gamma_M(\nu,\varrho)$ and any finite positive decomposition $\{x\}$ of the unity  
one has $|f({\mathbf 1})|\leq \sum_j|f(x_j)|=\sum_j|f(\sqrt{x_j}\sqrt{x_j})|\leq 
\sum_j \sqrt{\nu(x_j)\varrho(x_j)}$. According to {\sc Theorem 
\ref{genprob}\,(\ref{uhl})} therefore 
\begin{varequation}{\ensuremath{\star}}
\sqrt{P_M(\nu,\varrho)}\leq \inf_{\{x\}} \sum_j 
\sqrt{\nu(x_j)\varrho(x_j)}
\leq  \inf_{\{e\}} \sum_j 
\sqrt{\nu(e_j)\varrho(e_j)}
\end{varequation}
can be followed. That is, validity of (\ref{genprob.1b}) will imply that 
also (\ref{genprob.1a}) 
is true. To see that (\ref{genprob.1b}) holds, let $\varepsilon>0$. 
According to {\sc Theorem 
\ref{genprob}\,(\ref{genprob.1})} there exists invertible $x\in M_+$ and obeying 
$\nu(x)\varrho(x^{-1})<P_M(\nu,\varrho)+\varepsilon$. Since the map $y\,\longmapsto\,y^{-1}$ 
in restriction to the invertible elements of $M_+$ is normcontinuous, and since we are 
in a ${\mathsf W}^*$-algebra, in addition we may even suppose that $x$ satisfying the above estimate 
is chosen with finite spectrum, that is, $x=\sum_{j=1}^n \lambda_j e_j$ is fulfilled, 
with $\lambda_j>0$, and some finite decomposition $\{e_1,\ldots,e_n\}$ of the unity into 
mutually orthogonal orthoprojections of $M$. Using this spectral 
decomposition, one arrives at the expression 
$$
\nu(x)\varrho(x^{-1})=\sum_j \nu(e_j)\varrho(e_j)+ 
\sum_{j>k}\biggl\{\lambda_j\lambda_k^{-1} \nu(e_j)\varrho(e_k)+
\lambda_k\lambda_j^{-1} \nu(e_k)\varrho(e_j)\biggr\}.
$$       
Owing to strict positivity of $\lambda$'s and non-negativity 
of $\nu(e_j)$'s one has 
$$
\lambda_j\lambda_k^{-1} \nu(e_j)\varrho(e_k)+
\lambda_k\lambda_j^{-1} \nu(e_k)\varrho(e_j)\geq 2\,
\sqrt{\nu(e_j)\varrho(e_j)}\,\sqrt{\nu(e_k)\varrho(e_k)}\,,
$$ 
for each $j>k$. 
In fact, for $\sqrt{\nu(e_j)\varrho(e_j)}\sqrt{\nu(e_k)\varrho(e_k)}=0$ 
this is trivial, whereas 
in the other case the estimate follows from minimizing the positive function 
$F(t)=t\,\nu(e_j)\varrho(e_k)+ t^{-1}\,\nu(e_k)\varrho(e_j)$ over 
${\mathbb{R}}_+\backslash\{0\}$, which problem has a solution, since in this case both 
coefficients 
of $t$ and $t^{-1}$ are strictly positive. 
By means of this estimate and the above one finally arrives at    
\begin{varequation}{\ensuremath{\star\star}}
P_M(\nu,\varrho)+\varepsilon\geq\nu(x)\varrho(x^{-1})\geq
\biggl\{\sum_j \sqrt{\nu(e_j)\varrho(e_j)}
\biggr\}^2\,.
\end{varequation}
From this $\inf_{\{p\}} \sum_j \sqrt{\nu(p_j)\varrho(p_j)}\leq \sqrt{P_M(\nu,\varrho)+\varepsilon}$ 
is seen.  
Since $\varepsilon>0$ could have been chosen at will, 
$\sqrt{P_M(\nu,\varrho)}\geq \inf_{\{p\}} 
\sum_j \sqrt{\nu(p_j)\varrho(p_j)}$ follows, with 
$\{p\}$ extending over the finite decompositions of the unity into 
orthoprojections of $M$. From this and ($\star$) then (\ref{genprob.1a}) and 
(\ref{genprob.1b}) follow.

In order to prove (\ref{genprob.1c}), to given $\varepsilon>0$, for 
each $\delta>0$ by means 
of the decomposition $\{e_1,\ldots,e_n\}$ 
of the unity into 
orthoprojections $e_j$ obeying ($\star\star$) let 
us define a double system $\{y(\delta),x(\delta)\}\subset M$ by setting 
$x_j(\delta)=\mu_j(\delta)\,e_j$, $y_j(\delta)=\mu_j(\delta)^{-1}\, 
e_j$, with 
$$
\mu_j(\delta)=\sqrt[4]{\frac{\nu(e_j)+\delta}{\varrho(e_j)+\delta}}
$$
for each $j\leq n$. Then, also $\sum_j y_j^*(\delta)x_j(\delta)={\mathbf
1}$ holds, and therefore  
the double system $\{y(\delta),x(\delta)\}$ is a special case of those double systems 
considered in context of the infimum in (\ref{genprob.1c}). Hence, one has 
$\frac{1}{2}\inf_{\{y,x\}} 
\sum_j \bigl\{\nu(y_j^*y_j)+\varrho(x_j^*x_j)\bigr\}\leq F(\delta)$, for 
each $\delta>0$, with the auxiliary 
function $\delta\,\mapsto\,F(\delta)$ defined by $F(\delta)=\frac{1}{2}\sum_j 
\bigl\{\nu(y_j(\delta)^*y_j(\delta))+\varrho(x_j(\delta)^*x_j(\delta))
\bigr\}$. Since with this choice one easily infers that $F(\delta)$ may be expressed as  
\begin{eqnarray*}
F(\delta) & = & \sum_{j,\,\nu(e_j)\not=0} 
\frac{1}{2}\sqrt{\{\varrho(e_j)+\delta\}\nu(e_j)}\,
\sqrt{\frac{\nu(e_j)}{\nu(e_j)+\delta}} +\mbox{ }\\
& & \mbox{ }\hfill +\sum_{j,\,\varrho(e_j)\not=0} 
\frac{1}{2}\sqrt{\{\nu(e_j)+\delta\}\varrho(e_j)}\,
\sqrt{\frac{\varrho(e_j)}{\varrho(e_j)+\delta}}\,,
\end{eqnarray*}
in view of the previous and ($\star\star$) then  
\begin{varequation}{\ensuremath{\star^\prime}}
\lim_{\delta\to 0} F(\delta)=
\sum_j \sqrt{\nu(e_j)\varrho(e_j)}\leq 
\sqrt{P_M(\nu,\varrho)+\varepsilon}
\end{varequation}
can be followed. Therefore $\sqrt{P_M(\nu,\varrho)+\varepsilon}\geq 
\frac{1}{2}\inf_{\{y,x\}} 
\sum_j \bigl\{\nu(y_j^*y_j)+\varrho(x_j^*x_j)\bigr\}$ is seen. 
Since such procedure can be performed for each 
$\varepsilon>0$, one can
be assured that $\sqrt{P_M(\nu,\varrho)}\geq\frac{1}{2}\inf_{\{y,x\}} 
\sum_j \bigl\{\nu(y_j^*y_j)+\varrho(x_j^*x_j)\bigr\}$ is fulfilled, where 
$\{y,x\}$ is allowed to run through all finite double systems obeying 
$\sum_j y_j^*x_j={\mathbf
1}$. On the other hand, for each such double system and $f\in 
\Gamma_M(\nu,\varrho)$ one has 
$$
|f({\mathbf 1})|\leq \sum_j 
|f(y_j^*x_j)|\leq \sum_j 
\sqrt{{\nu(y_j^*y_j)\varrho(x_j^*x_j)}}\,.
$$
Now, for each two elements $x,y\in M$, from 
$\bigl\{\sqrt{\nu(y^*y)}-\sqrt{\varrho(x^*x)}\bigr\}^2\geq 0$ 
the estimate $\sqrt{\nu(y^*y)\varrho(x^*x)}\leq \frac{1}{2}\bigl\{
\nu(y^*y)+\varrho(x^*x)\bigr\}$ is inferred. Hence, the above 
estimate relating double systems can be continued accordingly and 
results in $|f({\mathbf 1})|\leq\frac{1}{2} \sum_j 
\bigl\{\nu(y_j^*y_j)+\varrho(x_j^*x_j)\bigr\}$. This has to hold for 
each $f\in \Gamma_M(\nu,\varrho)$ and finite double system $\{y,x\}$
obeying 
$\sum_j y_j^*x_j={\mathbf 1}$. Thus also 
$\sqrt{P_M(\nu,\varrho)}\leq\frac{1}{2}\inf_{\{y,x\}} \sum_j 
\bigl\{\nu(y_j^*y_j)+\varrho(x_j^*x_j)\bigr\}$ is seen. In view of the 
above then equality follows, that is, (\ref{genprob.1c}) is seen to hold. 
Note in context of ($\star^\prime$) that if an  
element $a(\delta)\in M$ is defined by means of the above $y_j(\delta)$ through the setting  
$a(\delta)=\sum_j y_j(\delta)^*y_j(\delta)$, one has $a(\delta)>{\mathbf 0}$, invertible with 
$a(\delta)^{-1}=\sum_j x_j(\delta)^*x_j(\delta)$, and then ($\star^\prime$) under the above premises 
on $\varepsilon$ equivalently also shows that  
$$
\lim_{\delta\to 0} \frac{1}{2}\,\bigl\{\nu((a(\delta))+\varrho(a(\delta)^{-1})
\bigr\}=\sum_j \sqrt{\nu(e_j)\varrho(e_j)}\leq 
\sqrt{P_M(\nu,\varrho)+\varepsilon}
$$
has to be fulfilled. 
Since $\varepsilon>0$ can be arbitrarily chosen, from the previous then even an 
estimate 
\begin{varequation}{\ensuremath{\star^{\prime\prime}}}
\sqrt{{P_M(\nu,\varrho)}}\geq 
\inf_{x>{\mathbf 0}}\frac{1}{2}\,\bigl\{\nu(x)+\varrho(x^{-1})\bigr\}
\end{varequation}
can be 
seen to be fulfilled, where now the infimum extends over all invertible, positive elements of $M$. On the other hand, 
for each invertible, positive element $x\in M$, one has the identity
\begin{subequation}[alph]\label{uhlprob}
\begin{equation}\label{uhlprob.1} 
\frac{1}{2}\,\biggl\{\sqrt{{\nu(x)}}-\sqrt{{\varrho(x^{-1})}}\biggr\}^2+
\sqrt{{\nu(x)\varrho(x^{-1})}}=\frac{1}{2}\,\bigl\{\nu(x)+\varrho(x^{-1})\bigr\}. 
\end{equation}
Taking the infimum over the 
invertible positive $x\in M$ on both sides and respecting non-negativity of  
$(1/2)\,\{\sqrt{{\nu(x)}}-\sqrt{{\varrho(x^{-1})}}\}^2$ then will show that the following estimate has to 
be fulfilled:
\begin{eqnarray}\label{uhlprob.2}
\inf_{x>{\mathbf 0}}\sqrt{{\nu(x)
\varrho(x^{-1})}} & \leq &  
\inf_{x>{\mathbf 0}}\frac{1}{2}\,\biggl\{\sqrt{\nu(x)}-\sqrt{\varrho(x^{-1})}\biggr\}^2
+ \inf_{x>{\mathbf 0}}\sqrt{{\nu(x)
\varrho(x^{-1})}}\nonumber \\
& \leq & \inf_{x>{\mathbf 0}}\frac{1}{2}\,\bigl\{\nu(x)+
\varrho(x^{-1})\bigr\}\,.
\end{eqnarray}
\end{subequation}
Hence, from {\sc Theorem \ref{genprob}\,(\ref{genprob.1})} one can 
conclude that $\sqrt{{P_M(\nu,\varrho)}}\leq  
\inf_{x>{\mathbf 0}}(1/2)\,\bigl\{\nu(x)+\varrho(x^{-1})\bigr\}$ has to hold. 
From this in view of 
($\star^{\prime\prime}$) the validity of (\ref{genprob.1u}) 
follows. 

Finally, for each 
$\varepsilon>0$ by the just proven (\ref{genprob.1u}) there exists an invertible 
$a>{\mathbf 0}$ obeying 
$\sqrt{P_M(\nu,\varrho)}+\varepsilon\geq (1/2)\,\bigl\{\nu(a)+\varrho(a^{-1})\bigr\}$.
In defining $y_\varepsilon=\sqrt{a}$ and $x_\varepsilon=\sqrt{a}^{\,-1}$ one has 
${\mathbf 1}=y_\varepsilon^*x_\varepsilon$, and the above estimate then turns into   
$(1/2)\,\bigl\{\nu(y_\varepsilon^*y_\varepsilon)+\varrho(
x_\varepsilon^*x_\varepsilon)\bigr\}\leq \sqrt{P_M(\nu,\varrho)}+\varepsilon$. 
On the other hand, according to (\ref{genprob.1c}) one has  
$$\sqrt{P_M(\nu,\varrho)}\leq\inf_{\{{\mathbf 1}=y^*x\}} 
(1/2)\bigl\{\nu(y^*y)+\varrho(x^*x)\bigr\}\leq (1/2)\bigl\{\nu(y_\varepsilon^*y_\varepsilon)+
\varrho(x_\varepsilon^*x_\varepsilon)\bigr\}\,.$$
From these estimates, and since $\varepsilon>0$ can be taken 
at will, validity of (\ref{genprob.1cc}) then gets evident. This completes the 
proof of all the assertions. 
\end{pf}

\subsection{Miscellaneous comments}\label{beweise.2}
In the following we will comment on the facts coming along with 
{\sc Corollary \ref{genprob}}, and will supplement them  
with further useful auxiliary results and remarks.

\subsubsection{Comments on {\sc Corollary \ref{genproba}\,(\ref{genprob.1a})--(\ref{genprob.1b})}
\textup{:} quadratic means}
\label{unter.1}
For normal states, $P_M(\nu,\varrho)$ is the same as the
generalized transition probability $T_M(\nu,\varrho)$ as
given in \cite{Cant:75}. 

The definition of V.\,Cantoni refers to the two 
probability measures $\nu(E_x(d\/\lambda))$ and 
$\varrho(E_x(d\/\lambda))$ over the Borel sets of ${\mathbb R}^1$ that 
can be naturally associated with two normal states $\nu,\,\varrho$ on 
$M$ through the projection valued measure $E_x(d\/\lambda)$ of a 
selfadjoint element, say $x\in M$, 
with spectral representation $x=\int_{{\mathbb R}^1} 
\lambda\,E_x(d\/\lambda)$ (remind that in a quantum mechanical context 
the hermitian elements are
the candidates of bounded observables). In line with some proposal 
of G.\,Mackey, see Chapter 2,\,2\,-\,2,\,2\,-\,6 in \cite{Mack:63}, 
and in accordance with some physically motivated axioms saying what 
properties of a 
`transition probability' should be considered as indispensable at all, 
see \cite{Miel:69,Gudd:78,Gudd:81} e.g., in \cite{Cant:75} one 
defines a generalized transition probability by 
\begin{equation}\label{cantoni}
T_M(\nu,\varrho)=\inf_{x\in M_{\mathrm{h}}} \biggl\{\int_{{\mathbb R}^1}
{\mathsf{QM}}_x(\nu,\varrho)(d\/\lambda)\biggr\}^2\,,
\end{equation} 
with the quadratic
means ${\mathsf{QM}}_x(\nu,\varrho)(d\/\lambda)=
\sqrt{\nu(E_x(d\/\lambda))\varrho(E_x(d\/\lambda))}$ of these measures, which is a Borel
measure on the line again. 
On carefully analyzing the quadratic means in the special case of 
two normal states and one of which is faithful at least, 
the proof that $P_M(\nu,\varrho)$ of {\em Definition \ref{genprob.2}} 
equals the expression (\ref{cantoni}) was given in \cite{ArRa:82}.
 
As has been yet remarked by S.\,Gudder, see {\em Theorem 
}\,1 in \cite{Gudd:81}, mathematically (\ref{cantoni}) amounts to
$\sqrt{T_M(\nu,\varrho)}=\inf_{\{e\}}\sum_j
\sqrt{\nu(e_j)\varrho(e_j)}$, which is (\ref{genprob.1b}) in this 
special case. 

In summarizing from all that, the news added 
through {\sc Corollary} \ref{genproba} to that subject are in the following\,:  
\begin{itemize}
\item[--] 
the expression 
in {\sc Corollary} \ref{genproba}\,(\ref{genprob.1b}) reflects those 
aspects behind (\ref{cantoni}) 
which remain valid for {\em any} positive linear forms (not only normal 
ones) on a ${\mathsf W}^*$-algebra; 
\item[--]
the expression in {\sc Corollary} \ref{genproba}\,(\ref{genprob.1a}) can be taken as 
the common general 
${\mathsf C}^*$-algebraic essence of the matter around quadratic means.    
\end{itemize}
 
\subsubsection{Comments on {\sc Corollary \ref{genproba}\,(\ref{genprob.1c})--(\ref{genprob.1cc})}
\textup{:} some seminorms on $M$}
\label{unter.2}  
For normal states, (\ref{genprob.1c}) had been conjectured by D.\,Buchholz, 
motivated by some application to relativistic quantum field 
theory, and has been proved in the special case of ${\mathsf B}({\mathcal H})$ 
in \cite{Buch:90}, see eq.\,(2.10) there. 

But note that there the intention was to deal even with certain vector states of some 
$^*$-algebras of (unbounded) operators. 
In contrast to this, in the following we will strictly adhere to the (bounded) context of a 
${\mathsf W}^*$-algebra $M$ and positive linear forms. 

To start discussions  
around {\sc Corollary \ref{genproba}\,(\ref{genprob.1c})--(\ref{genprob.1cc})}, for given 
$\nu,\varrho\in M_+^*$ let us consider two realvalued functions on $M$, $\tau_{\nu,\varrho}$ and 
$\upsilon_{\nu,\varrho}$, which are defined at $z\in M$ by   
\begin{subequation}\label{buch}
\begin{eqnarray}\label{buch.1}
\tau_{\nu,\varrho}(z) & = & \inf_{\{y,x\}\subset M,\,z=\sum_{j\leq n} 
y_j^*x_j} 
\frac{1}{2}\sum_j\biggl\{\nu\bigl(y_j^*y_j\bigr)+\varrho\bigl(
x_j^*x_j\bigr)\biggr\}\,,\\ 
\label{uhlbuch}
\upsilon_{\nu,\varrho}(z) & = & \ \ \ \ \ \ \ \ \ \inf_{z= y^*x}\frac{1}{2}\,\bigl\{\nu(y^*y)+\varrho(
x^*x)\bigr\}\,.
\end{eqnarray} 
Thereby, within the former expression the infimum is to be taken over all finite 
double systems $\{y,x\}$ of operators of $M$ obeying $z=\sum_{j\leq 
n} y_j^*x_j$, with $n\in{\mathbb{N}}$ arbitrarily chosen. 
For notational simplicity subsequently use the shortcut notation $z=\{y,x\}$ 
whenever such type of relation occurs. If we want to consider only 
minimal systems of that kind ($n=1$), 
to which e.g~within (\ref{uhlbuch}) is referred to, the 
condition $z=y^*x$ will be explicitely used. 

Note that the assertions of {\sc Corollary \ref{genproba}\,(\ref{genprob.1c})--(\ref{genprob.1cc})} 
then read as  
\begin{equation}\label{tau1}
\upsilon_{\nu,\varrho}({\mathbf 1})=\tau_{\nu,\varrho}({\mathbf 1})=\sqrt{{P_M(\nu,\varrho)}}\,.
\end{equation} 
Also it is obvious from the structure of the expression within definition 
(\ref{buch.1}) that 
$\tau_{\nu,\varrho}$ is a seminorm, whereas from (\ref{uhlbuch}) it is obvious that 
$\tau_{\nu,\varrho}$ is a lower bound for $\upsilon_{\nu,\varrho}$\,:
\begin{equation}\label{taubound}
\tau_{\nu,\varrho}(z)\leq
\upsilon_{\nu,\varrho}(z)\,.
\end{equation}
Remark that in relativistic quantum field 
theory there was some hope that seminorms of $\tau$-type 
should be useful in proving existence of non-trivial (weak) intertwiners between so-called 
standard representations \cite{Yngv:73,Buch:90}, 
these standard representations roughly corresponding to the cyclic 
$^*$-representations of $\nu$ and $\varrho$ accordingly, in our bounded 
context (for the context see 
also \cite{Haag:92}, especially Definition 2.2.14). Clearly, this in a 
specific setting is the (highly non-trivial) analog over unbounded observable 
algebras of the (comparably trivial) task of 
analyzing the structure of the set $\Gamma_M(\nu,\varrho)$ in the bounded 
case. In the bounded case, the above idea reduces to the inquiry  
for upper bounds of $f\in \Gamma_M(\nu,\varrho)$ which read in terms of 
the seminorm $\tau_{\nu,\varrho}$, that is, one is looking for estimates by $\tau_{\nu,\varrho}$ from above
\begin{equation}\label{ub}
\forall\,z\in M\,:\ \ |f(z)|\leq c\,\tau_{\nu,\varrho}(z)\,,
\end{equation}
for some real constant $c>0$, for instance.
 
More precisely, the news around 
{\sc Corollary \ref{genproba}\,(\ref{genprob.1c})--(\ref{genprob.1cc})} 
to be annotated by our comments will be in the following\,:
\begin{itemize}
\item[--] 
with respect to the seminorm (\ref{buch.1}) the estimate (\ref{ub}) holds, with $c=1$, 
and this estimate being the 
best possible in favor of the above task, that is, $\Gamma_M(\nu,\varrho)$ appears to be 
trivial, $\Gamma_M(\nu,\varrho)=\{0\}$, 
if, and only if, $\tau_{\nu,\varrho}$ is trivial, $\tau_{\nu,\varrho}\equiv 0$; 
\item[--]
the seminorm $\tau_{\nu,\varrho}$ can be calculated exactly even if    
$\{y,x\}$ under the infimum in (\ref{buch.1}) is bent to be varied only 
through minimal double systems with $z=y^*x$, that 
is, according to this and (\ref{uhlbuch}) one has 
$\tau_{\nu,\varrho}=\upsilon_{\nu,\varrho}$ to hold;
\item[--]
when seen in form of (\ref{tau1}), in generalizing from {\sc Corollary \ref{genproba}\,(\ref{genprob.1c})} for each $\nu,\varrho\in M_+^*$ and given $z\in M$ an (heuristic useful) interpretation of 
the values of the seminorm $\tau_{\nu,\varrho}$ in terms of `transition probability' (and 
thus in terms of the Bures-distance) between certain 
inner derived from $\{\nu,\varrho\}$ positive linear forms can be given. 
\end{itemize}
It is plain to see that the answers to the corresponding items can be read off 
as straightforward consequences from the following result\,:
\begin{corolla}\label{buch0}
For each $a,b\in M$ and $z=a^*b$ the following holds\,:
\begin{eqnarray}\label{buch.2}
\tau_{\nu,\varrho}(z)=\upsilon_{\nu,\varrho}(z) & = & \sup_{f\in 
\Gamma_M(\nu,\varrho)} |f(z)|=\nonumber \\ 
& = & \sqrt{P_M\bigl(\nu,\varrho^z\bigr)}=\sqrt{P_M\bigl(\nu^a,\varrho^b\bigr)}\,.
\end{eqnarray} 
\end{corolla}
\end{subequation}
\medskip
\begin{pf}
First note that each finite double system $\{y,x\}$ obeying
${\mathbf 1}=\{y,x\}$ through setting $\tilde{y}_j=y_ja$ and
$\tilde{x}_j=x_jb$, respectively, provides another finite double system of the same length 
$\{\tilde{y},\tilde{x}\}$ with 
$a^*b=\{\tilde{y},\tilde{x}\}$ (especially, minimal double systems will be transformed 
into minimal ones again). Hence, in view of 
{\sc Corollary \ref{genproba}\,(\ref{genprob.1c})--(\ref{genprob.1cc})} and (\ref{buch.1})--(\ref{uhlbuch}) one can conclude as follows: 
\begin{eqnarray*}
\sqrt{{P_M\bigl(\nu^a,\varrho^b\bigr)}} & = & (1/2)\inf_{{\mathbf 1}=\{y,x\}} 
\sum_j\bigl\{\nu^a\bigl(y_j^*y_j\bigr)+\varrho^b\bigl(
x_j^*x_j\bigr)\bigr\}\\
& = & (1/2)\inf_{{\mathbf 1}=\{y,x\}} 
\sum_j\bigl\{\nu\bigl(\tilde{y}_j^*\tilde{y}_j\bigr)+\varrho\bigl(
\tilde{x}_j^*\tilde{x}_j\bigr)\bigr\}\\
& \geq &  
(1/2)\inf_{a^*b=\{y,x\}} 
\sum_j\bigl\{\nu\bigl(y_j^*y_j\bigr)+\varrho\bigl(
x_j^*x_j\bigr)\bigr\}\\
& = & \tau_{\nu,\varrho}(a^*b)\,.
\end{eqnarray*} 
Thus, the following estimate has been established:
\begin{varequation}{\ensuremath{\circ}}
\tau_{\nu,\varrho}(a^*b)\leq \sqrt{P_M\bigl(\nu^a,\varrho^b\bigr)}\,.
\end{varequation}
Also, if to the pair $\{\nu,\varrho\}$ a representation $\{\pi,{\mathcal 
K}\}$ as in the premises of {\sc Theorem \ref{genprob}\,(\ref{genprob.3})} 
is chosen, with fixed $\varphi\in {\mathcal S}_{\pi,M}(\nu)$ and 
$\psi\in {\mathcal S}_{\pi,M}(\varrho)$ then   
obviously also $\pi(a)\varphi\in {\mathcal S}_{\pi,M}(\nu^a)$ and 
$\pi(b)\psi\in {\mathcal S}_{\pi,M}(\varrho^b)$ are fulfilled. Application of 
(\ref{schief0}) with respect to $\{\nu,\varrho\}$, $\{\nu^a,\varrho^b\}$ and 
$\{\nu,\varrho^z\}$ will yield that $\langle\pi(\cdot)k\psi,\varphi\rangle$, 
$\langle\pi(\cdot)k\pi(b)\psi,\pi(a)\varphi\rangle$ and 
$\langle\pi(\cdot)k\pi(z)\psi,\varphi\rangle$, 
respectively, will be running through all of $\Gamma_M(\nu,\varrho)$, $\Gamma_M(\nu^a,\varrho^b)$ and 
$\Gamma_M(\nu,\varrho^z)$, respectively, if $k$ is supposed to be varied through 
all of $(\pi(M)^{\,\prime})_1$. Now, for each $k\in(\pi(M)^{\,\prime})_1$ one has 
$\langle k\pi(b)\psi,\pi(a)\varphi\rangle=\langle k\pi(z)\psi,\varphi\rangle=
\langle\pi(z)k\psi,\varphi\rangle$. Hence, in line with {\sc Theorem \ref{genprob}\,(\ref{uhl})},  
when the latter accordingly is applied to these three special situations 
at hand, under the premise of $z=a^*b$ the estimate 
($\circ$) can be continued as follows\,: 
\begin{varequation}{\ensuremath{\circ^\prime}}
\tau_{\nu,\varrho}(z)\leq 
\sqrt{P_M\bigl(\nu^a,\varrho^b\bigr)}=\sqrt{P_M\bigl(\nu,\varrho^z\bigr)}=
\sup_{f\in 
\Gamma_M(\nu,\varrho)} |f(z)|\,.
\end{varequation}
Now, suppose $z=\{y,x\}$ in context of $\{\nu,\varrho\}$. By  
definition of $\Gamma_M(\nu,\varrho)$, for $f\in 
\Gamma_M(\nu,\varrho)$ one has 
\begin{eqnarray*}
|f(z)| & \leq & \sum_j 
|f\bigl(y_j^*x_j\bigr)|\leq \sum_j 
\sqrt{{\nu\bigl(y_j^*y_j\bigr)\,\varrho\bigl(x_j^*x_j\bigr)}
}\\
& \leq & \frac{1}{2}\sum_j\bigl\{\nu\bigl(y_j^*y_j\bigr)+
\varrho\bigl(x_j^*x_j\bigr)\bigr\}\,.
\end{eqnarray*}
From this in view of (\ref{buch.1}) 
${\sup_{f\in 
\Gamma_M(\nu,\varrho)} }|f(z)|\leq \tau_{\nu,\varrho}(z)$ follows, which with the help of 
(\ref{taubound}) can be turned into
\begin{varequation}{\ensuremath{\circ^{\prime\prime}}}
{\sup_{f\in 
\Gamma_M(\nu,\varrho)} }|f(z)|\leq \tau_{\nu,\varrho}(z)\leq \upsilon_{\nu,\varrho}(z)\,.
\end{varequation} 
On the other hand, for 
$\varepsilon>0$, {\sc Corollary \ref{genprob}\,(\ref{genprob.1cc})} can be applied to the pair 
$\{\nu,\varrho^z\}$ and yields invertible $a>{\mathbf 0}$ obeying 
$\sqrt{{P_M(\nu,\varrho^z)}}+\varepsilon\geq (1/2)\,\bigl\{\nu(a)+\varrho^z(a^{-1})\bigr\}$.
Let us define $y=\sqrt{a}$ and $x=\sqrt{{a}}^{\,-1}\,z$. Then, $z=y^*x$ and 
$\bigl\{\nu(a)+\varrho^z(a^{-1})\bigr\}=\bigl\{\nu(y^*y)+\varrho(
x^*x)\bigr\}$ are fulfilled. Hence, in view of above $\upsilon_{\nu,\varrho}(z)\leq 
\sqrt{{P_M(\nu,\varrho^z)}}+\varepsilon$ can be followed. Since $\varepsilon>0$ can be taken 
at will, from the latter in accordance with (\ref{uhlbuch}) we get $
\upsilon_{\nu,\varrho}(z)\leq \sqrt{P_M(\nu,\varrho^z)}$.
Upon taking together this  
with ($\circ^{\prime\prime}$) and ($\circ^{\prime}$) we can conclude that in fact 
equality has to occur within ($\circ^{\prime\prime}$) and ($\circ^{\prime}$), that is, (\ref{buch.2}) holds. This closes the proof of {\sc Corollary \ref{buch0}}.  
\end{pf}
\subparagraph{Proof \textup{(of {\sc Theorem \ref{buch00}})}.}
Formula {\sc Theorem \ref{buch00}}\,(\ref{buch.20}) is given by one of the particular 
subequations coming along with (\ref{buch.2}). Moreover, according to another subequation of 
the latter $\upsilon_{\nu,\varrho}(z)=P_M(\nu^a,\varrho^b)^{1/2}$ holds. Inserting 
this into (\ref{pcont.1}) in view of (\ref{uhlbuch}) yields
$d_{\mathsf B}(\nu^a,\varrho^b)^2=\nu(a^*a)+\varrho(b^*b)-\inf_{z=y^*x}
\bigl\{\nu(y^*y)+\varrho(x^*x)\bigr\}
=\sup_{z=y^*x}\bigl\{\nu(a^*a-y^*y)+\varrho(b^*b-x^*x)\bigr\}$, and 
which is {\sc Theorem \ref{buch00}}\,(\ref{uhlbuch.1}).
 
\begin{remark}\label{uhlrem}
\begin{enumerate}
\item\label{pcont.3}
Without proof remark that $P_M(\nu,\varrho)=0$ is equivalent with 
$\nu\perp\varrho$, see e.g.~\cite{AlUh:84}. Recall that 
orthogonality of two ${\mathsf C}^*$-algebraic positive linear forms $\nu,\,\varrho$ 
is defined as $\|\nu-\varrho\|_1=\|\nu\|_1 + \|\varrho\|_1$.  
\item\label{uhlrem.1}
Especially for states $\nu,\varrho$ occuring along with quantum physical 
problems over an algebra of observables $M$, one is inclined to give 
$P_M(\nu,\varrho)$ a (quantum) probabilistic interpretation. {\sc Corollary \ref{buch0}} in such context 
will tell us that an interpretation 
which reads in terms of the transition probability, but now between the 
`perturbed' states $\nu^a$ and $\varrho^b$, extends also to the value of 
the rather abstractly defined seminorms $M\ni z\,\longmapsto\,\tau_{\nu,\varrho}(z)$ at $z=a^*b$. 
Thus, if to given pair $\{\nu,\varrho\}$ of states and in accordance with (\ref{buch.2}) and the previous 
item (\ref{pcont.3})   
those operators $a$,\,$b$ are considered which are solutions 
of the equation $\tau_{\nu,\varrho}(a^*b)=0$ (and for which both $\nu^a$ and $\varrho^b$ are 
states again), then these might be interpreted as all possible elementary `operations' 
(i.e.~inner implementable perturbations) 
driving $\{\nu,\varrho\}$ into mutually orthogonal states.
\item\label{uhlrem.3}
Due to the mentioned interpretation of the values of the 
seminorm $\tau_{\nu,\varrho}$ in terms of $\sqrt{P_M}$, which manifests itself by   
(\ref{buch.2}), some subadditivity property of $\sqrt{P_M}$ in respect to inner derived positive linear 
forms can be followed\,:
$$
a^*b=\sum_{j\leq n} a_j^*b_j\ \ \Longrightarrow\ \ 
\sqrt{P_M\bigl(\nu^a,\varrho^b\bigr)}\leq \sum_{j\leq n}\sqrt{P_M\bigl(\nu^{a_j},
\varrho^{b_j}\bigr)}\,.
$$
\item\label{uhlrem.1u}
The fact that $\tau_{\nu,\varrho}=\upsilon_{\nu,\varrho}$ holds  
is mainly due to our restriction to {\em bounded} operator algebras and  
cannot be expected to extend simply to a context with $^*$-algebras 
of unbounded operators.    
\end{enumerate}
\end{remark}
\subsubsection{Comments on {\sc Corollary \ref{genproba}\,(\ref{genprob.1u})}\textup{:} 
minimizing abelian algebras}
\label{unter.3}
That {\sc Corollary \ref{genproba}\,(\ref{genprob.1u})} is a notable 
result on its own rights - and is not something to be easily abandoned - has been 
recognized only recently, and as such 
will be discussed here (and more detailed in the next section) for the 
first time. 

In comparing the item in question to 
{\sc Theorem \ref{genprob}}\,(\ref{genprob.1}) one at once notices that the essential 
difference with the latter result lies in the fact that  
under the infimum instead of a geometrical means now the arithmetical 
means of the same two expressions enters.   

Quite naturally, in context of {\sc Corollary \ref{genproba}}\,(\ref{genprob.1u}) 
(and in context of  
{\sc Theorem \ref{genprob}}\,(\ref{genprob.1}) as well) a main interest will be in describing 
the structure of those invertible $x\in M_+$ at which by the expression of 
$\frac{1}{2}\,\bigl\{\nu(x)+\varrho(x^{-1})\bigr\}$ (or  
$\sqrt{\nu(x)\,\varrho(x^{-1})}$, respectively) the  
(common) infimum $\sqrt{P_M(\nu,\varrho)}$ is nearly attained. 
Such problems and related questions  
we are going to discuss now. Thereby, for these purposes of estimation theory 
{\sc Corollary \ref{genproba}}\,(\ref{genprob.1u}) seems to be 
better suited than {\sc Theorem \ref{genproba}}\,(\ref{genprob.1}). 
For instance, the map $x\,\longmapsto\, 
\frac{1}{2}\,\bigl\{\nu(x)+\varrho(x^{-1})\bigr\}$ is more sensitive 
to certain variations of the positive invertible operator $x\in M$ than the map 
$x\,\longmapsto\, 
\sqrt{\nu(x)\,\varrho(x^{-1})}$ is (compare the behavior of both under the change 
$x\,\mapsto\,\lambda\,x$, for real $\lambda>0$, simply).

Relating the quality of the mentioned approximation one has the 
following simple facts (cf.~also Theorem 4.4 in \cite{Albe:92.1}). 
\begin{corolla}\label{eqval}
Let $\nu,\varrho\in M^*_+$, and be $\{x\}\subset M_+$ a sequence of 
invertible elements. The following facts are equivalent\textup{:}
\begin{enumerate}[0000]
\item\label{eqval.1}
$\sqrt{P_M(\nu,\varrho)}=\lim_{n\to\infty} \frac{1}{2}
\,\bigl\{\nu\bigl(x_n\bigr)+\varrho\bigl(x_n^{-1}\bigr)\bigr\}
\,;$ 
\item\label{eqval.2}
$\sqrt{P_M(\nu,\varrho)}=\lim_{n\to\infty} \nu\bigl(x_n\bigr)=\lim_{n\to\infty} 
\varrho\bigl(x_n^{-1}\bigr)\,.$
\end{enumerate}
Moreover, if ${\mathrm{Comm}}[M]$ is the family of all abelian ${\mathsf W}^*$-subalgebras of $M$ 
with the same unity as $M$,  
then one has 
\begin{enumerate}[0000]
\setcounter{enumi}{2}
\item\label{eqval.3}
$P_M(\nu,\varrho)=\inf_{R\in {\mathrm{Comm}}[M]} P_R\bigl(\nu|_R,\varrho|_R\bigr)\,.$
\end{enumerate} 
\end{corolla} 
\medskip
\begin{pf}
In view of eqs.~(\ref{uhlprob}) the asserted equivalence immediately follows from 
{\sc Theorem \ref{genproba}}\,(\ref{genprob.1}) and 
{\sc Corollary \ref{genproba}}\,(\ref{genprob.1u}). Also (\ref{eqval.3}) can 
be seen as an obvious consequence from each of these items.  
\end{pf}   
Now, for given pair $\{\nu,\varrho\}$ of positive linear forms 
a set ${\mathsf{Min}}_M(\nu,\varrho)$ will be defined as follows\,:
$$
{\mathsf{Min}}_M(\nu,\varrho)=\biggl\{x\in M_+:\ \sqrt{P_M(\nu,\varrho)}=
\frac{1}{2}\,\bigl\{\nu(x)+\varrho(x^{-1})\bigr\}\biggr\}\,.
$$
The elements of ${\mathsf{Min}}_M(\nu,\varrho)$ will be called {\em minimizing} 
(positive invertible) elements to the pair $\{\nu,\varrho\}$, where in this notation 
tacitely to the context with {\sc Corollary \ref{genproba}}\,(\ref{genprob.1u}) is referred to. 

Note that since the set of all invertible positive elements is neither 
compact nor closed, it is a non-trivial problem to 
decide from a concrete pair $\{\nu,\varrho\}$ of positive linear forms 
whether or not the infimum 
within {\sc Corollary \ref{genproba}}\,(\ref{genprob.1u}) is a minimum. 

In fact, in general this cannot happen, as the following simple counterexample shows. 
\begin{exam}\label{gegen}
According to elementary spectral theory for invertible $y\in M_+$ one has 
$y\geq \|y^{-1}\|^{-1}{\mathbf 1}$. 
Hence, for each pair $\{\nu,\varrho\}\not=\{0,0\}$ of positive linear forms and  
for each invertible $x\in M_+$ one infers that
$\{\nu(x)+\varrho(x^{-1})\}/2\geq 
\{{\|\nu\|_1}/{\|x^{-1}\|}+{\|\varrho\|_1}/{\|x\|}\}/2>0$ 
has to be fulfilled. On the other hand, according to 
{\em Remark \ref{uhlrem}}\,(\ref{pcont.3}),  
in the special case of $\nu\perp\varrho$ one has $\sqrt{P_M(\nu,\varrho)}=0$. Thus, 
in view of the previous estimate in case of a nontrivial pair of mutually 
orthogonal positive linear forms ${\mathsf{Min}}_M(\nu,\varrho)=\emptyset$ 
holds.  
\end{exam}
On the other hand, there exist also classes where this question can be 
answered affirmatively. A criterion relating this matter  
is easily obtained 
from {\sc Corollary \ref{eqval}}\,(\ref{eqval.1})--(\ref{eqval.2}) and reads as follows\,:
\begin{equation}\label{eqvalb}
x\in M_+,\,\sqrt{P_M(\nu,\varrho)}=\nu(x)=\varrho\bigl(x^{-1})\  
\Longleftrightarrow\ x \in {\mathsf{Min}}_M(\nu,\varrho)
\,.
\end{equation}
\begin{exam}\label{commb}
Suppose $\varrho=\nu^a$, with $a\in M_+$ being {\em invertible}. Then, in view of  
{\sc Theorem \ref{uhlrem.2}} the criterion (\ref{eqvalb}) gets applicable with 
$x=a$ and yields 
that the infimum in 
{\sc Corollary \ref{genproba}}\,(\ref{genprob.1u}) is a minimum. 
\end{exam}
Let us refer to an abelian ${\mathsf W}^*$-subalgebra 
$R\subset M$ with ${\mathbf 1}\in R$ as {\em minimizing abelian subalgebra} if  
the infimum within {\sc Corollary \ref{eqval}}\,(\ref{eqval.3}) is a minimum and is attained 
at $R$. For instance, if ${\mathsf{Min}}_M(\nu,\varrho)\not=\emptyset$ is fulfilled then   
in line with the above the infimum is attained at each   
subalgebra $R$ which is generated by ${\mathbf 1}$ and some particular 
$x\in{\mathsf{Min}}_M(\nu,\varrho)$. Thus, in generalizing from the problem on existence of minimizing elements 
the more general question on existence of minimizing abelian subalgebras naturally arises.

\setcounter{equation}{0}
\section{Special subjects}\label{minimal}   
\subsection{Minimizing elements}\label{unter.31} 
In this paragraph we inquire for existence and uniqueness 
of minimizing positive invertible elements, and we are going to derive some 
results on the structure of ${\mathsf{Min}}_M(\nu,\varrho)$.
Let $x,z\in M_+$ be any two  
invertible positive elements. Let $\delta=(z-x)$. Then, the following algebraic identity 
can be easily 
checked to hold:
\begin{subequation}\label{inveq}
\begin{equation}\label{inveq.1}
z^{-1}=x^{-1}-x^{-1}\delta x^{-1}+\Delta(z,x)\,,
\end{equation}
where $\Delta(z,x)=m(z,x)^*m(z,x)$ holds, and $m(z,x)$ is defined by 
\begin{varequation}{\ensuremath{\star}}
m(z,x)=(x^{-1/2}\delta x^{-1/2})(x^{-1/2}z x^{-1/2})^{-1/2}x^{-1/2}\,.
\end{varequation}
By construction of $\Delta(z,x)$ and by invertibility of $z,x$ from 
($\star$) then   
\begin{equation}\label{inveq.3}
\Delta(z,x)\in M_+,\mbox{ {\em with} }\,\biggl\{
\Delta(z,x)={\mathbf 0}\ \Longleftrightarrow\ 
\delta={\mathbf 0}\biggr\}
\end{equation}
can be followed. Also, since $x^{-1/2}\delta x^{-1/2}$ is commuting 
with $x^{-1/2}z x^{-1/2}$, from ($\star$) yet another expression for $m(z,x)$ 
can be obtained, and which reads as
\begin{equation}\label{inveq.2}  
m(z,x)=(x^{-1/2}z x^{-1/2})^{-1/2}x^{-1/2}\delta x^{-1}\,.
\end{equation}
With the help of (\ref{inveq.1}) and in the previous notations one then finds
\begin{eqnarray}\label{inveq.4}
\frac{1}{2}\biggl\{\nu(z) & + & \varrho(z^{-1})\biggr\}  - 
\frac{1}{2}\biggl\{\nu(x)+\varrho(x^{-1})\biggr\}=\nonumber \\
& & \frac{1}{2}\biggl\{
\nu(\delta)-\varrho(x^{-1}\delta x^{-1})\biggr\}+\frac{1}{2}\varrho(\Delta(z,x))\,.
\end{eqnarray}
\end{subequation}
Note that the set $M_+^{\mathsf{inv}}$ of all invertible positive elements of 
$M$ is an open non-pointed 
subcone within the real Banach space $\{M_{\mathrm{h}},\|\cdot\|\}$ of the hermitian portion 
of $M$. Hence, for a particular $x\in M_+^{\mathsf{inv}}$ and given $y\in M_{\mathrm{h}}$, for 
all $t\in {\mathbb R}$ sufficiently small $z_t=x+ty\in M_+^{\mathsf{inv}}$ has to hold (one might take $|t|<\|x^{-1}yx^{-1}\|^{-1}$, e.g.). In this special situation the formula 
(\ref{inveq.4}) at such parameter $t$ reads as
\begin{eqnarray}\label{diff}
\frac{1}{2}\biggl\{\nu(z_t) & + & \varrho(z_t^{-1})\biggr\}  - 
\frac{1}{2}\biggl\{\nu(x)+\varrho(x^{-1})\biggr\}=\nonumber \\
& & \frac{t}{2}\biggl\{
\nu(y)-\varrho(x^{-1}y x^{-1})\biggr\}+\frac{t^2}{2}\varrho(\Delta_t(y|x))\,,
\end{eqnarray}
where $\Delta_t(y|x)=t^{-2}\Delta(z_t,x)$ is defined for $t\not=0$, and 
at $t=0$ we let $\Delta_0(y|x)=\|\cdot\|-\lim_{t\to 0}t^{-2}\Delta(z_t,x)=x^{-1}y x^{-1}y x^{-1}$. 
We are now ready for the following redefinition of ${\mathsf{Min}}_M(\nu,\varrho)$. 
\begin{satz}\label{minimisator}
For any $\nu,\varrho\in M_+^*$ the following holds\textup{:}
\begin{equation}\label{mini}
{\mathsf{Min}}_M(\nu,\varrho)=\biggl\{x\in M_+^{\mathsf{inv}}:\ \nu(y)=\varrho(x^{-1}y x^{-1})
,\,\forall\,y\in M_{\mathrm{h}}\biggr\}\,. 
\end{equation}
\end{satz}
\medskip
\begin{pf}
Suppose $x\in {\mathsf{Min}}_M(\nu,\varrho)$. Then, for each fixed $y\in M_{\mathrm{h}}$, and 
all $t\in {\mathbb R}\backslash \{0\}$ sufficiently small, in accordance with (\ref{diff}) 
$$-\biggl|
\nu(y)-\varrho(x^{-1}y x^{-1})\biggr|\geq -|t|\,\varrho(\Delta_t(y|x))$$ has to 
hold. Having in mind that 
according to the above $t\,\mapsto\,\Delta_t(y|x)$ is norm-continuous at $t=0$, one then has 
$\lim_{t\to 0} |t|\,\varrho(\Delta_t(y|x))=0$. 
In view of the previous estimate from this  
$\nu(y)=\varrho(x^{-1}y x^{-1})$ is seen. 

On the other hand, assume $x\in M_+^{\mathsf{inv}}$ such 
that, for each $y\in M_{\mathrm{h}}$, $\nu(y)=\varrho(x^{-1}y x^{-1})$ is satisfied. 
For each other $z\in M_+^{\mathsf{inv}}$, let $\delta=(z-x)=y$. One then 
especially has $\{\nu(\delta)-\varrho(x^{-1}\delta x^{-1})\}=0$. Hence, (\ref{inveq.4}) 
can be applied and owing to positivity of $\Delta(z,x)$ and $\varrho$ 
yields $\frac{1}{2}\bigl\{\nu(z)+\varrho(z^{-1})\bigr\}  - 
\frac{1}{2}\bigl\{\nu(x)+\varrho(x^{-1})\bigr\}\geq 0$. Hence, since $z$ can be 
arbitrarily chosen from $M_+^{\mathsf{inv}}$, $x\in {\mathsf{Min}}_M(\nu,\varrho)$ follows. 
This completes the proof of (\ref{mini}).
\end{pf}
After these preliminaries we may now summarize as follows. 
\begin{theorem}\label{stabi}
Let $M$ be a ${\mathsf W}^*$-algebra. For $\nu,\varrho\in M_+^*$ one has
\textup{:}
\begin{enumerate}[0000]
\item\label{stabi.1}
$\;\,\ {\mathsf{Min}}_M(\nu,\varrho)\not=\emptyset\ \Longleftrightarrow\ \exists\,a 
\in M_+^{\mathsf{inv}}:\,\varrho=\nu^a\,;$
\item\label{stabi.1a}
$\;\,\ {\mathsf{Min}}_M(\nu,\varrho)=
\bigl\{x+I_\nu\bigr\}\bigcap M_+^{\mathsf{inv}},\,\forall\,x\in 
{\mathsf{Min}}_M(\nu,\varrho)\,;$
\item\label{stabi.2}
$\# {\mathsf{Min}}_M(\nu,\varrho)=1\ \Longleftrightarrow\ \exists\,a 
\in M_+^{\mathsf{inv}}:\varrho=\nu^a,\,\nu\textup{\ \em is faithful }.$
\end{enumerate}  
\end{theorem}
\medskip
\begin{pf}
According to {\em Example \ref{commb}}, for $\varrho=\nu^a$ with $a \in M_+^{\mathsf{inv}}$ one has 
$a\in {\mathsf{Min}}_M(\nu,\varrho)$. On the other hand, if ${\mathsf{Min}}_M(\nu,\varrho)\not=\emptyset$ 
is supposed, in line with formula (\ref{mini}) and since linear forms on a ${\mathsf C}^*$-algebra are 
uniquely determined through their values on the hermitian portion, $\nu=\varrho(x^{-1}(\cdot)x^{-1})$ has 
to be fulfilled, for some $x \in M_+^{\mathsf{inv}}$. That is, $\varrho=\nu^a$ holds, with $a=x$. 
In summarizing, (\ref{stabi.1}) is valid.

To see (\ref{stabi.1a}), suppose $x\in {\mathsf{Min}}_M(\nu,\varrho)$ and 
be $z\in M_+^{\mathsf{inv}}$. According to the previous then $\varrho=\nu^x$, 
and therefore from (\ref{mini}) and (\ref{inveq.4}) one infers that 
$z\in {\mathsf{Min}}_M(\nu,\varrho)$ happens if, and only if, 
$\nu(x\Delta(z,x)x)=0$ is fulfilled. By construction of $\Delta(z,x)$ 
the latter is equivalent with $m(z,x)x\in I_\nu$, see (\ref{ide}). 
According to (\ref{inveq.2}) 
the latter is the same as $(x^{-1/2}z x^{-1/2})^{-1/2}x^{-1/2}\delta
\in I_\nu$, with $\delta=(z-x)$. Since $I_\nu$ is a left ideal and 
$(x^{-1/2}z x^{-1/2})^{-1/2}x^{-1/2}$ is invertible, from this we 
finally conclude that for $z\in M_+^{\mathsf{inv}}$ the condition 
$z\in {\mathsf{Min}}_M(\nu,\varrho)$ has to be equivalent with $\delta\in I_\nu$. 
Owing to ${\mathsf{Min}}_M(\nu,\varrho)\subset M_+^{\mathsf{inv}}$ 
this is (\ref{stabi.1a}).

In order to see (\ref{stabi.2}), remark first that for faithful $\nu$ 
one has $I_\nu=\{{\mathbf 0}\}$. Hence, 
from the just proved (\ref{stabi.1a}) uniqueness of a minimizing element 
evidently follows. On the other hand, for an eventually existing 
$r\in I_\nu\backslash \{{\mathbf 0}\}$ owing to $r^*r=|r|^2$ 
also $|r|\in I_\nu\backslash \{{\mathbf 0}\}$ follows, see (\ref{ide}). 
Hence, since $a\in {\mathsf{Min}}_M(\nu,\varrho)$ is invertible, 
by standard facts and owing to $z\geq a$ also 
$z \in M_+^{\mathsf{inv}}$ follows, for $z=a+|r|$.   
By (\ref{stabi.1a}) this however then implies  
$z\in {\mathsf{Min}}_M(\nu,\varrho)$. Since $z\not=a$ holds we therefore have $\#{\mathsf{Min}}_M(\nu,\varrho)>1$, 
for non-faithful $\nu$. Taking together 
this with the previous yields (\ref{stabi.2}).     
\end{pf} 
Since ${\mathsf{Min}}_M(\varrho,\nu)=\bigl\{x^{-1}:x\in 
{\mathsf{Min}}_M(\nu,\varrho)\bigr\}$ holds, from 
{\sc Theorem \ref{stabi}\,(\ref{stabi.1a})} for 
${\mathsf{Min}}_M(\nu,\varrho)\not=\emptyset$ one infers that 
both positive linear forms have to be 
faithful or not, only simultaneously. In reversing this another class of 
counterexamples is easily obtained.    
\begin{exam}\label{stabiex}
Let $\nu,\varrho\in M_+^*$. Suppose exactly one of the two forms to be faithful. Then, 
the infimum in {\sc Corollary \ref{genproba}\,(\ref{genprob.1u})} cannot be attained 
on the invertible positive elements of $M$. 
\end{exam}
\begin{remark}\label{minstand}
According to {\sc Theorem \ref{stabi}\,(\ref{stabi.1})} minimizing 
elements can exist if, and only if, each one of the two positive linear forms of a 
pair $\{\nu,\varrho\}$ can be inner derived by means of some positive invertible 
element from the other one in quest. That is, in fact 
all these cases are yet covered by {\em Example \ref{commb}}. 
\end{remark} 
As announced at the end of \ref{unter.3}, 
the next best question to be raised is to inquire for existence of 
a commutative ${\mathsf W}^*$-subalgebra $R$ of 
$M$, with ${\mathbf 1}\in R$, such that the infimum in 
{\sc Corollary \ref{eqval}}\,(\ref{eqval.3}) could be attained. 

\subsection{Minimizing commutative subalgebras}\label{unter.32} 
Start with examples where minimizing abelian subalgebras exist but which 
are found slighly beyond of {\em Example \ref{commb}}.  
\begin{exam}\label{comma}
Suppose $\varrho=\nu^a$, for $a\in M_+$. By functional calculus (use the spectral representation 
theorem within the ${\mathsf W}^*$-algebra $M$) one infers $a(a+\varepsilon\,{\mathbf 1})^{-1}a\leq a$ 
to hold, 
for each real $\varepsilon>0$. Hence,  
$a_\varepsilon=(a+\varepsilon\,{\mathbf 1})\in M_+^{\mathsf{inv}}$ with  
$\varrho(a_\varepsilon^{-1})\leq \nu(a)$.  
Owing to this, {\sc Theorem \ref{genproba}}\,(\ref{genprob.1}) and 
{\sc Theorem \ref{uhlrem.2}} (or (\ref{au76a}), equivalently) then 
$\nu(a)=\sqrt{P_M(\nu,\varrho)}\leq \nu(a_\varepsilon)=\nu(a)+\varepsilon\,\|\nu\|_1$ as well as  
$\nu(a)^2=P_M(\nu,\varrho)\leq \varrho(a_\varepsilon^{-1})\,\nu(a_\varepsilon)\leq
\nu(a)^2+\varepsilon\,\nu(a)\|\nu\|_1$ 
are obtained. Upon performing the limit $\varepsilon\to 0$ 
in both relations and regarding 
{\sc Corollary \ref{eqval}}\,(\ref{eqval.1})--(\ref{eqval.2}) 
will give that the ${\mathsf W}^*$-subalgebra generated by 
$a$ and ${\mathbf 1}$ can be chosen 
as minimizing commutative subalgebra $R$. 
\end{exam}
The fact that a subalgebra $R$ be minimizing for a given pair $\{\nu,\varrho\}$ 
implies that some very specific additional conditions have to be fulfilled. 
An important instance of such conditions occurs in context of those minimizing 
subalgebras which come along with {\em Example \ref{comma}}.
\begin{lemma}\label{opt}
Suppose $\nu,\varrho\in M_+^*$ and let $R$ be 
a ${\mathsf W}^*$-subalgebra of $M$ such that $\varrho|_R=(\nu|_R)^a$ holds, 
for some $a\in R_+$. 
Then, whenever $R$ is minimizing for $\{\nu,\varrho\}$ the relation
\begin{equation}\label{proj.-1}
\nu^p(a)-\nu^{p^\perp}(a)=\nu(a)
\end{equation}
holds, for each orthoprojection $p\in M$ obeying $p^\perp\in I_\varrho$.
\end{lemma}
\smallskip
\begin{pf}
Let $P=P_M(\nu,\varrho)$. The assumption that $R$ be minimizing together with the reasoning of  
{\em Example \ref{comma}}\, when applied in respect of  
$\{\nu|_R,\varrho|_R\}$ over $R$ prove that, for  
$a_\varepsilon=a+\varepsilon\,{\mathbf 1}$ with $\varepsilon>0$, one  
has $\sqrt{P}=\nu(a)=\lim_{\varepsilon\to 0} \nu(a_\varepsilon)=
\lim_{\varepsilon\to 0} \varrho(a_\varepsilon^{-1})$. 
Now, let $u=p+\lambda\,p^\perp$, 
with real $\lambda\not=0$. Define 
$a_\varepsilon(\lambda)=u^*a_\varepsilon u$. 
Then, for each $\varepsilon >0$ one has $a_\varepsilon(\lambda)\in M_+^{\mathsf{inv}}$. Note also 
that the assumption on $p$ saying that $p^\perp\in I_\varrho$ be fulfilled 
together with the special structure of $u$ imply 
$\varrho(y)=\varrho(pyp)=\varrho(u^*y u)=\varrho(u^{-1} y u^{{-1}*})$ to be 
fulfilled, 
for each $y\in M$. Hence, by construction of $a_\varepsilon(\lambda)$ then especially 
also $\lim_{\varepsilon\to 0}\varrho(a_\varepsilon(\lambda)^{-1})=
\lim_{\varepsilon\to 0}\varrho(a_\varepsilon^{-1})=\sqrt{P}$ follows. On the other hand, since   
$\nu(a_\varepsilon(\lambda))=\nu^p(a_\varepsilon)+
2\lambda\, \Re\,\nu(p^\perp a_\varepsilon p) +
\lambda^2 \nu^{p^\perp}(a_\varepsilon)$ is fulfilled, in view of the above one arrives at 
$\lim_{\varepsilon\to 0}\nu(a_\varepsilon(\lambda))=\nu^p(a)+
2\lambda\, \Re\,\nu(p^\perp a p) +
\lambda^2 \nu^{p^\perp}(a)$. 
Note that according to {\sc Theorem \ref{genprob} (\ref{genprob.1})} the estimate  
$\lim_{\varepsilon\to 0}\nu(a_\varepsilon(\lambda))\,\varrho(a_\varepsilon(\lambda)^{-1})\geq P$ has to 
be fulfilled, which condition in view of the previous amounts to requiring 
$\sqrt{P}\,\{\nu^p(a)+
2\lambda\, \Re\,\nu(p^\perp a p) +
\lambda^2 \nu^{p^\perp}(a)\}\geq P=\sqrt{P}\,
\{\nu^p(a)+
2\,\Re\{\nu(p^\perp a p)\} +\nu^{p^\perp}(a)\}$, for all reals $\lambda\not=0$. 
That is, 
$$
2\sqrt{P}(\lambda-1)\biggl\{\Re\,\nu(p^\perp a p)+
\frac{1}{2}(\lambda+1)\,\nu^{p^\perp}(a)
\biggr\}\geq 0
$$
has to be fulfilled, for each real $\lambda\not=0$.\\
Suppose $P\not=0$ first. In considering the previous estimate 
for $\lambda>1$ one infers that    
$\Re\,\nu(p^\perp a p)+\frac{1}{2}(\lambda+1)\,
\nu^{p^\perp}(a)\geq 0$ has to be fulfilled, whereas 
for $\lambda<1$ we see that $\Re\,\nu(p^\perp a p)
+\frac{1}{2}(\lambda+1)\,
\nu^{p^\perp}(a)\leq 0$ has to be fulfilled. 
Upon performing the limits $\lambda \searrow 1$ and 
$\lambda \nearrow 1$ within the mentioned relations for $\lambda>1$ and $\lambda<1$, respectively, 
and then comparing the results will show that 
$\nu^{p^\perp}(a)=-\Re\,\nu(p^\perp a p)$ 
has to be fulfilled. By means of this then $\nu(a)=
\nu^p(a)+2\,\Re\,\nu(p^\perp a p)+\nu^{p^\perp}(a)=\nu^p(a)-\nu^{p^\perp}(a)$ is seen. 
This proves the result in case of $P\not=0$. \\
Finally, for $P=0$ one has $\nu(a)=0$. 
Owing to $a\geq {\mathbf 0}$ then $a\in I_\nu$. 
Hence also $0=\nu(p a)=\nu(a p)$, and therefore from  
$\nu^{p^\perp}(a)=\nu(a)-2\,\Re\,\nu(a p)+\nu^p(a)$ one then gets 
$\nu^p(a)-\nu^{p^\perp}(a)=0$ which is in accordance with 
(\ref{proj.-1}) in this special case.   
\end{pf}
Having in mind {\em Example \ref{stabiex}}\, remark that for faithful $\nu$ and $\varrho=\nu^a$, 
with $a\in M_+$ and $\ker a\not=\{\mathbf 0\}$, 
the most simple situations arise where  
{\em Example \ref{comma}}\, provides cases which go beyond of {\em Example \ref{commb}}. 
Less trivial situations of that kind arise from generalizing      
{\em Example \ref{rad}}\, and modifying those  
arguments, along which we have been following within {\em Example \ref{comma}}. 
The result in question, which however will be proved here only in some sketchy way, reads as follows. 
\begin{satz}\label{eqvala}
Let $\{\nu,\varrho\}$, with {\em normal}  
$\nu,\varrho\in M_+^*$, and support orthoprojections 
which are mutually $\leq$-comparable, say $s(\varrho)\leq s(\nu)$ be fulfilled.   
Then a minimizing commutative ${\mathsf W}^*$-subalgebra $R$ of $M$ exists. 
\end{satz}
\subparagraph{Sketch of proof.}
Remark first that for {\em normal} 
positive linear forms $\nu,\varrho$ with supports obeying 
$s(\varrho)\leq s(\nu)$ the problem 
in quest via some appropriately chosen {\em normal} $^*$-representation $\{\pi,{\mathcal K}\}$, 
which obeys ${\mathcal S}_{\pi,M}(\nu)\not=\emptyset$ and 
${\mathcal S}_{\pi,M}(\varrho)\not=\emptyset$, always can be reduced to the 
analogous problem over the $vN$-algebra 
$N=\pi(M)^{\,\prime\prime}$. In this setting, with given 
$\varphi\in {\mathcal S}_{\pi,M}(\nu)$, the assumption 
about the supports can be shown to ensure existence of some (possibly 
unbounded) selfadjoint positive linear operator $A$, which is affiliated 
with $N$ and which obeys $\psi=A\varphi\in {\mathcal S}_{\pi,M}(\varrho)$. 
Note that since $A$ is affiliated with $N$, the operator $A$ 
can be chosen to be independent from the particularly chosen $\varphi$ 
within ${\mathcal S}_{\pi,M}(\nu)$. Let $\nu_\pi$ and 
$\varrho_\pi$ be the vector functionals generated by $\varphi$ and $\psi$ 
over the $vN$-algebra $N$. Extending the notion  
`inner derived positive linear form' slighly to include at least also such 
situations with vector forms on $N$ and (unbounded) 
positive selfadjoint linear operators 
affiliated with $N$, for $\varrho_\pi=\nu_\pi^A$ one 
easily proves that formula (\ref{au76a}) remains true in the sense 
of $\sqrt{P_N(\nu_\pi,\varrho_\pi)}=\nu_\pi(A)=
\langle A\varphi,\varphi\rangle$. Since then also the arguments 
raised in context of {\em Example \ref{comma}} are easily justified to 
remain valid with $A_\varepsilon=A+\varepsilon\,{\mathbf 1}$ instead of 
$a_\varepsilon$, following along the same line of conclusions as in 
{\em Example \ref{comma}} will provide $P_N(\nu_\pi,\varrho_\pi)=
P_R(\nu_\pi|_R,\varrho_\pi|_R)$, with $R$ being the commutative 
$vN$-subalgebra of $N$ generated by the spectral resolution of $A$. 
Finally, since always  
$P_N(\nu_\pi,\varrho_\pi)=P_M(\nu,\varrho)$ is fulfilled (note that $\nu_\pi\circ\pi=\nu$ and 
$\varrho_\pi\circ\pi=\varrho$ hold), 
in view of {\em normality} of $\pi$, which implies that 
even $N=\pi(M)$ holds, the just mentioned result about $\nu_\pi$,\,$\varrho_\pi$ over $N$     
can be rewritten easily into one over $M$.

\subsection{Least minimizing commutative subalgebra}\label{unter.33}
\subsubsection{Generalities on the problem}\label{unter.331}
It is plain to see (from each of the items of {\sc Corollary \ref{genproba}}, e.g.) that 
the map $R\,\longmapsto\,\sqrt{P_R(\nu|_R,\varrho|_R)}$, $\nu,\varrho\in M_+^*$,  
with respect to the inclusion $\subset$ between ${\mathsf W}^*$-subalgebras of 
$M$ behaves $\leq$-(anti-)monotoneous. Hence, if there is a minimizing commutative subalgebra $R$, then also each larger than this   
commutative subalgebra has to be minimizing. 

Going the other way around in this context is less trivial. For instance, one might ask 
for existence of a {\em least} minimizing commutative 
${\mathsf W}^*$-subalgebra of 
$M$ with the same unit. In case of existence of a least minimizing subalgebra the latter 
will be denoted by ${\mathcal R}_M(\nu,\varrho)$. 

Note that a least minimizing subalgebra must not exist in either case of a pair $\{\nu,\varrho\}$ where 
a minimizing commutative subalgebra exists. 
To formulate a result on this, 
for the following agree to make use of $R[x]$ as notation for the commutative 
${\mathsf W}^*$-subalgebra of $M$ which is generated by ${\mathbf 1}$ and the hermitian element 
$x\in M_{\mathrm h}$. Then, the simplest counterexamples against existence of a least 
minimizing algebra can be generated along the following auxiliary construction.
\begin{lemma}\label{neg}
Suppose $\varrho=\nu^x$ holds, with $x\in M_+$. Then, for each  
$k\in I_\nu\cap M_+$, $R[x+k]$ is a minimizing abelian subalgebra 
to $\{\nu,\varrho\}$. 
In case      
\begin{equation}\label{neg.1}
\varrho\not\in {\mathbb R}_+\,\nu\,,\ \bigcap_{k\in I_\nu\cap M_+} R[x+k]={\mathbb C}\cdot{\mathbf 1}
\end{equation} 
is fulfilled, there cannot exist a least element among all 
minimizing commutative subalgebras to the pair $\{\nu,\varrho\}$. 
\end{lemma}
\smallskip
\begin{pf}
It is easily inferred from (\ref{cauchy}) and (\ref{ide}) that also 
$\varrho=\nu^{x+k}$ holds, for each $k\in I_\nu\cap M_+$.  
By {\em Example \ref{comma}}\, then $R[x+k]$ will be a 
special minimizing commutative subalgebra. 
Also, from {\em Definition \ref{genprob.2}}\, with the help of known properties of the 
Cauchy-Schwarz inequality one easily infers that for each pair $\{\nu,\varrho\}$ of 
positive linear forms $\sqrt{P_M(\nu,\varrho)}\leq \sqrt{\|\nu\|_1\|\varrho\|_1}$ is fulfilled, 
with equality occuring if, and only if, $\varrho=\lambda\cdot\nu$ happens for some 
non-negative real 
$\lambda$. On the other hand, from the structure of 
{\sc Corollary \ref{genproba}\,(\ref{genprob.1u})} it is 
easily seen that $\sqrt{P_M(\nu,\varrho)}=\sqrt{\|\nu\|_1\|\varrho\|_1}$ is equivalent with 
the fact that ${\mathbb C}\cdot{\mathbf 1}$ be among the minimizing subalgebras. 
Now, assume $\nu,\varrho$ as in (\ref{neg.1}). Then, according to the first of 
the previously mentioned 
facts the second condition in (\ref{neg.1}) in case of existence of a 
least minimizing subalgebra implied the latter to be trivial, 
whereas by the first condition in (\ref{neg.1}) and owing to the second of 
the above mentioned facts the trivial algebra ${\mathbb C}\cdot{\mathbf 1}$ is excluded 
from being a 
minimizing subalgebra. Thus, a least minimizing subalgebra cannot 
exist in this case. 
\end{pf}
Unfortunately, the condition (\ref{neg.1}) can be satisfied easily, e.g.~it can be shown to be 
fulfilled for any two non-commuting pure states (the following $2\times 2$\,-\,case exemplarily 
can stand for any situation of this kind; we omit the details).  
\begin{exam}\label{neg.2}
Let $M={\mathsf M}_2({\mathbb C})$ be the full algebra of $2\times 2$-matrices with complex 
entries, $p,q\in M$ one-dimensional orthoprojections, with 
$[p,q]=pq-qp\not={\mathbf 0}$. Let $x=p+\varepsilon\,p^\perp$, with $0<\varepsilon<1$, and be 
$\nu\in M_+^*\backslash \{0\}$ with $\nu(q)=0$ (such positive linear form trivially exists). 
Define $\varrho=\nu^x$. Then, $q\in I_\nu\cap M_+$, and in line with 
the first part of {\sc Lemma \ref{neg}} for both $x$ and $y=x+q$ one has that 
$R[x]$ and $R[y]$ are minimizing commutative subalgebras, which owing to the assumptions 
obey $[x,y]\not={\mathbf 0}$, and therefore both have to be non-trivial as well as  
cannot be the same, $R[x]\not=R[y]$. Since each non-trivial commutative subalgebra of 
${\mathsf M}_2({\mathbb C})$ can be generated by exactly two atoms, from the previous 
then $R[x]\cap R[y]={\mathbb C}\cdot{\mathbf 1}$ has to be followed. This especially means that 
condition (\ref{neg.1}) is fulfilled, and thus in accordance with the other assertion of  
{\sc Lemma \ref{neg}} a least minimizing subalgebra cannot exist. 
\end{exam}
The above negative result and the previous counterexample together 
with some view on the structure 
of the condition 
(\ref{neg.1}) indicate that existence of a least minimizing abelian subalgebra  
seems to depend sensitively 
from the size as well as from the mutual position of the 
kernel-ideals $I_\nu$ and $I_\varrho$ 
to each other (cf. also {\sc Lemma \ref{opt}}). Remind that the kernel-ideal $I_\nu$ in a ${\mathsf W}^*$-algebra gets 
manageable especially if $\nu$ is supposed to be {\em normal}. 
In this case $I_\nu=M s(\nu)^\perp$ 
holds, where $s(\nu)$ is the support orthoprojection of the normal positive 
linear form $\nu$ (be careful about the context; the same notation $s(x)$ will be also used for the 
support of an hermitian element 
$x\in M_{\mathrm{h}}$, which subsequently also will play a r\^{o}le). 
Unfortunately, even in the normal case only very few answers are 
known on this subject, except we are in the special case with $\varrho\ll \nu$ which relates to 
{\em Example \ref{rad}}, and where sufficiently many examples of minimizing abelian subalgebras are known. 
Before going into the details some auxiliary notion relating a general  
pair $\{\nu,\varrho\}$ of {\em normal} positive linear forms will be introduced\,: 
\begin{defi}\label{proj}
Let $R\subset M$ be a ${\mathsf W}^*$-subalgebra of $M$, which contains the unity of $M$. 
$R$ is called $\{\nu,\varrho\}$-{\em projective} provided the 
condition
\begin{equation}\label{proj.0}
\forall y\in R\,:\ \ \nu^{s(\varrho)}(y)=\nu(ys(\varrho))
\end{equation}
is fulfilled ($R$ will be simply referred to as 
{\em projective subalgebra} if the ordered pair is unambiguously given by the context). 
\end{defi}
\begin{exam}\label{zentral}
For a normal positive linear form $\nu$ the unital subalgebra 
$M^\nu$ defined by $M^\nu=\{x\in M:\nu(xy)=\nu(yx),\,\forall\,y\in M\}$ is a 
${\mathsf W}^*$-subalgebra of $M$, which usually is called $\nu$-centralizer. 
Obviously, if the support $s(\varrho)$ of another 
normal positive linear form $\varrho$ obeys $s(\varrho)\in M^\nu$, then  
relation (\ref{proj.0}) is automatically fulfilled, 
for each ${\mathsf W}^*$-subalgebra $R$ of $M$. Hence, in this case each 
such $R$ is $\{\nu,\varrho\}$-projective.
\end{exam}
\begin{remark}\label{projrem}
\begin{enumerate}
\item\label{proj.2}
Since for each normal positive linear form $\nu$ one has $s(\nu)\in M^\nu$, 
according to {\em Example \ref{zentral}} in case of normal $\nu,\varrho\in M_+^*$ with equal supports, 
$s(\nu)=s(\varrho)$, each subalgebra $R$ of $M$ is both $\{\nu,\varrho\}$-   
and $\{\varrho,\nu\}$-projective. 
\item\label{proj.3}
Obviously, for given $\{\nu,\varrho\}$ the set of all $\{\nu,\varrho\}$-projective subalgebras of $M$ is non-void, 
and each 
subalgebra of a projective subalgebra is projective again. Also, the set of all projective 
subalgebras of $M$ is closed with respect to intersections. 
\item\label{proj.4}
Suppose $\varrho=\nu^x$, for a pair $\{\nu,\varrho\}$ of normal positive linear forms, with 
$x\in M_+$ obeying $x s(\varrho)=s(\varrho) x$. Then, according to {\em Example \ref{comma}} and since then 
obviously (\ref{proj.0}) is fulfilled for $R=R[x]$, the latter subalgebra is an  
example of a {\em minimizing abelian projective subalgebra} of $M$ for $\{\nu,\varrho\}$. 
\item\label{proj.5}
Suppose under the conditions of the previous (\ref{proj.4}) that a least minimizing abelian 
subalgebra ${\mathcal R}_M(\nu,\varrho)$ exists. According to the previous 
two items it follows that ${\mathcal R}_M(\nu,\varrho)$ has to be projective, too.   
\end{enumerate}
\end{remark}
\subsubsection{Radon-Nikodym theorem and minimizing projective subalgebras}\label{unter.332}
For the following recall that in case of   
$\varrho\ll\nu$ the Radon-Nikodym operator 
$x=\sqrt{d\/\varrho/d\/\nu}$ of $\varrho$ relative to 
$\nu$ is understood to be the unique element $x\in M_+$ which obeys both $\varrho=\nu^x$ and 
$s(x)\leq s(\nu)$.   
\begin{lemma}\label{aux0}
Suppose $\nu,\varrho\in M_+^*$ are normal, with $\varrho\ll\nu$. 
Let $R$ be any minimizing abelian projective subalgebra of M  
for $\{\nu,\varrho\}$. Then, the following facts are valid\textup{:}
\begin{enumerate}[0000] 
\item\label{aux0.1}
$\forall\,k\in s(\nu)^\perp M_+ s(\nu)^\perp:\, 
R[\sqrt{d\/\varrho/d\/\nu}+k]$ is minimizing, projective\,;
\item\label{aux0.1a}
$\exists\,k\in s(\nu)^\perp M_+ s(\nu)^\perp:\, 
R[\sqrt{d\/\varrho/d\/\nu}+k]\subset R\,.$
\end{enumerate}
\end{lemma}
\smallskip  
\begin{pf}
According to {\em Example \ref{rad}}\, and {\em Example \ref{comma}}\, 
one knows that the assumptions ensure that minimizing abelian 
subalgebras in fact have to exist. Since $\nu$ is normal, as mentioned 
above $I_\nu=M s(\nu)^\perp$ holds. 
Hence $I_\nu\cap M_+=s(\nu)^\perp M_+ s(\nu)^\perp$ holds, and then by 
{\sc Lemma \ref{neg}} we know that  
the formula in (\ref{aux0.1}) provides minimizing abelian subalgebras. Moreover, since 
$\varrho\ll \nu$ implies $s(\sqrt{d\/\varrho/d\/\nu})=s(\varrho)\leq s(\nu)$, one obviously 
has that each of $\sqrt{d\/\varrho/d\/\nu}+k$, with $k\in s(\nu)^\perp M_+ s(\nu)^\perp$, commutes 
with $s(\varrho)$. Hence, by {\em Remark \ref{projrem}} (\ref{proj.4}) all the subalgebras 
given in accordance with (\ref{aux0.1}) also are projective. 
Thus, it remains to be shown that each minimizing 
abelian projective subalgebra $R$ has a subalgebras as given in line with (\ref{aux0.1}). 
Note that for $\varrho=0$ the assertion holds since then ${\mathbb C}\cdot {\mathbf 1}$ is minimizing. 
In line with this, we are going to prove the previous assertion in the non-trivial case with $\nu,\varrho\not=0$.

Let $R$ be any minimizing abelian projective subalgebra to the 
given pair $\{\nu,\varrho\}$. Note that by their very 
definitions the conditions of normality for 
a positive linear form, as well as the relation $\ll$ among normal 
positive linear forms, are 
hereditary conditions when considered in restriction to  
${\mathsf W}^*$-subalgebras of $M$. Thus 
especially we also find $\varrho|_R\ll \nu|_R$ on $R$. 
Therefore we have unique Radon-Nikodym operators 
$x=\sqrt{d\/\varrho/d\/\nu}$ and $z=\sqrt{d\/\varrho|_R/d\/\nu|_R}$. As mentioned above we 
then especially have $s(x)=s(\varrho)\leq s(\nu)$, and since 
$\varrho\not=0$ is supposed in this case, we also have $z\not={\mathbf 0}$. 
The assumption that $R$ be minimizing together with the reasoning of  
{\em Example \ref{comma}}\, when applied for  
$\{\nu,\varrho\}$ over $M$, and for 
$\{\nu|_R,\varrho|_R\}$ over $R$, respectively, prove that for   
$x_\varepsilon=x+\varepsilon\,{\mathbf 1}$ and 
$z_\varepsilon=z+\varepsilon\,{\mathbf 1}$, with $\varepsilon>0$, one  
has $\lim_{\varepsilon\to 0} \nu(x_\varepsilon)=\lim_{\varepsilon\to 0} 
\varrho(x_\varepsilon^{-1})=\nu(x)=\sqrt{P_M(\nu,\varrho)}=
\sqrt{P_R(\nu|_R,\varrho|_R)}=\nu(z)=\lim_{\varepsilon\to 0} \nu(z_\varepsilon)=
\lim_{\varepsilon\to 0} \varrho(z_\varepsilon^{-1})$. Hence, since 
$\delta=(z_\varepsilon-x_\varepsilon)=(z-x)$ and $\varrho=\nu^x$ hold, 
upon taking the limit $\varepsilon\to 0$ within the relations which occur if 
(\ref{inveq.4}) is considered for $z_\varepsilon,x_\varepsilon$ 
instead of $z,x$ we will arrive at 
\begin{subequation}\label{radcomma}
\begin{equation}\label{radcomm.2}
0=-\lim_{\varepsilon\to 0} \nu(x x_\varepsilon^{-1}\delta x_\varepsilon^{-1} x)+
\lim_{\varepsilon\to 0} \nu(x x_\varepsilon^{-1}\delta z_\varepsilon^{-1}
\delta x_\varepsilon^{-1}x)\,,
\end{equation}
where also the special form of $m(z_\varepsilon,x_\varepsilon)$ arising 
along with (\ref{inveq.2}) has been taken into account. Also note that by 
elementary facts on spectral theory  
$s_\varepsilon=x x_\varepsilon^{-1}=x_\varepsilon^{-1}x$ is positive 
for each $\varepsilon$. Also, if positive 
reals are regarded as a directed set in its descending ordering, then $\{s_\varepsilon\}\subset M_+$ 
turns into an  ascendingly directed net of positive elements of $M$, with 
$s_\varepsilon\leq s(x)$, and which has 
the support orthoprojection $s(x)$ of $x$ as least upper bound, 
that is, ${\mathrm{l.u.b.}}\{s_\varepsilon: \varepsilon>0\}=s(x)$ 
is fulfilled. In passing note 
that the assertion on monotonicity can be understood as a 
special consequence of the fact saying that 
the function ${\mathbb R}_+\backslash\{0\}\ni t\,
\mapsto\,t^{-1}$ is {\em operator-(anti)monotoneous} over 
$M_+^{\mathsf{inv}}$ (for generalities on that and related, 
see \cite{BeSh:55,Dono:74}). Since $s(x)=s(\varrho)$ holds, from the previous with the help of 
(\ref{cauchy}) for each $y\in M$ 
one then easily concludes that $|\nu^{s(\varrho)}(y)-\nu(s_\varepsilon y s_\varepsilon)|
\leq 
|\nu^{s(\varrho)}(y)-\nu(y s_\varepsilon)|+
|\nu(y s_\varepsilon)-\nu(s_\varepsilon y s_\varepsilon)|\leq 
2\,\|y\|\,\sqrt{\nu(s(x)-s_\varepsilon)\,\|\nu\|_1}$ must be fulfilled. 
From this owing to normality of $\nu$ and 
${\mathrm{l.u.b.}}\{s_\varepsilon: \varepsilon>0\}=s(x)=s(\varrho)$   
\begin{equation}\label{radcomm.2a}
\forall\,y\in M\,:\ \nu^{s(\varrho)}(y)=\lim_{\varepsilon\to 0} 
\nu(s_\varepsilon y s_\varepsilon)
\end{equation}  
follows. From this in view of (\ref{radcomm.2}) especially also follows that both limits within 
(\ref{radcomm.2}) really exist. 
Now, remember that $R$ by assumption is both minimizing and projective. 
Hence, in view of {\sc Lemma \ref{opt}} and {\em Definition \ref{proj}} 
both, (\ref{proj.-1}) with $a=z$ and $p=s(\varrho)$, as well as the particular 
case of the relation in (\ref{proj.0}) at $y=z$ hold. That is, $\nu(z)=
\nu^{s(\varrho)}(z)-\nu^{s(\varrho)^\perp}(z)$ and $\nu(s(\varrho)^\perp z s(\varrho))=0$ are 
fulfilled. From the latter $\nu(z)=
\nu^{s(\varrho)}(z)+2\,\Re\,\nu(s(\varrho)^\perp z s(\varrho))+\nu^{s(\varrho)^\perp}(z)=
\nu^{s(\varrho)}(z)+\nu^{s(\varrho)^\perp}(z)$ is obtained. This together with the former 
provides the following relation\,: 
\begin{equation}\label{rel}
\nu^{s(\varrho)}(z)=\nu(z)\,.
\end{equation} 
But then, since owing to $s(x)=s(\varrho)$ 
also $\nu^{s(\varrho)}(x)=\nu(x)$ must be fulfilled, 
$\nu^{s(\varrho)}(\delta)=\nu(\delta)$ can be followed. 
Remind that $\nu(\delta)=0$ holds.   
In specializing $y=\delta$ within (\ref{radcomm.2a}), in line of  
the previous (\ref{radcomm.2}) can be also read as   
\begin{equation}\label{radcomm.3}
\lim_{\varepsilon\to 0} \nu(s_\varepsilon\delta z_\varepsilon^{-1}
\delta s_\varepsilon)=0\,.
\end{equation}  
\end{subequation}
Also note that by the estimate $z_\varepsilon\leq (\|z\|+\varepsilon)\,
{\mathbf 1}$, which is valid by triviality,  
$(\|z\|+\varepsilon)^{-1}\,{\mathbf 1}\leq z_\varepsilon^{-1}$ is implied.  
But then, since the linear map $M\ni y\,\mapsto\,s_\varepsilon\delta y
\delta s_\varepsilon\in M$ is positive, from the previous and by  
positivity of $\nu$ one infers   
$\nu(s_\varepsilon\delta z_\varepsilon^{-1}
\delta s_\varepsilon)\geq (\|z\|+\varepsilon)^{-1}
\nu(s_\varepsilon\delta^2 s_\varepsilon)\geq 0$.  
Regarding the limit of the latter as $\varepsilon\to 0$, and respecting that 
$\|z\|\not=0$ holds, in view of    
(\ref{radcomm.3}) yields $\nu^{s(\varrho)}(\delta^2)=0$, finally.   
Owing to $s(\varrho)\leq s(\nu)$ from this  
$\delta s(\varrho)={\mathbf 0}$ follows. Hence, since $s(\varrho)=s(x)$ and 
$z\in R_+\subset M_+$ hold, the conclusion is that $z=x+k$ has to be 
fulfilled, with $k=z s(\varrho)^\perp=s(\varrho)^\perp z\in 
s(\varrho)^\perp M_+ s(\varrho)^\perp$. But note that by $\nu(\delta)=0$ then also 
$\nu(k)=0$ follows. By positivity of $k$ and $ks(\varrho)^\perp=k$ from this we conclude to 
$s(\nu)s(\varrho)^\perp k 
s(\varrho)^\perp s(\nu)={\mathbf 0}$, which is equivalent with 
$k s(\varrho)^\perp s(\nu)={\mathbf 0}$, and thus $k$ even 
must obey  
$k\in s(\nu)^\perp M_+ s(\nu)^\perp$. This together with the obvious relation 
$R[x+k]=R[z]\subset R$ is the assertion of (\ref{aux0.1a}).  
\end{pf}
\begin{theorem}\label{radcomm}
Suppose $\varrho\ll \nu$ is fulfilled, for 
normal positive linear forms $\nu,\varrho\in M_+^*$, with faithful $\nu$.  
The following facts hold\textup{:}
\begin{enumerate}[0000]
\item\label{radcomm1}
provided ${\mathcal R}_M(\nu,\varrho)$ exists it obeys 
\begin{equation}\label{radcomm.1}
{\mathcal R}_M(\nu,\varrho)=R\biggl[\sqrt{d\/\varrho/d\/\nu}\biggr]\,; 
\end{equation}
\item\label{radcomm2}
if also $\varrho$ is faithful then ${\mathcal R}_M(\nu,\varrho)$ exists.
\end{enumerate}
\end{theorem}
\smallskip
\begin{pf}
By {\sc Lemma \ref{aux0}}\,(\ref{aux0.1}) one knows that 
$R=R\bigl[\sqrt{d\/\varrho/d\/\nu}\bigr]$ 
is minimizing and projective. Hence, if ${\mathcal R}_M(\nu,\varrho)$ is assumed to exist then by 
{\em Remark \ref{projrem}} (\ref{proj.4})--(\ref{proj.5}) the minimizing subalgebra  
${\mathcal R}_M(\nu,\varrho)\subset R\bigl[\sqrt{d\/\varrho/d\/\nu}\bigr]$ has to be also projective 
(occasionally remark that this conclusion does not rely on the premise on faithfulness of $\nu$).   
Hence, {\sc Lemma \ref{aux0}}\,(\ref{aux0.1a}) can be applied to  
$R={\mathcal R}_M(\nu,\varrho)$. By faithfulness of $\nu$ one has $s(\nu)^\perp={\mathbf 0}$ and then the mentioned  
application yields $R\subset R\bigl[\sqrt{d\/\varrho/d\/\nu}\bigr]$, and in view of the above the 
formula (\ref{radcomm.1}) then is seen to hold, that is, (\ref{radcomm1}) is valid. 
To see (\ref{radcomm2}), note that in this case ${\mathbf 1}=s(\nu)=s(\varrho)$ holds, 
which via {\em Remark \ref{projrem}} (\ref{proj.2}) 
implies that  
{\sc Lemma \ref{aux0}}\,(\ref{aux0.1a}) can be applied to {\em each} minimizing $R$. In line with this 
then $R\bigl[\sqrt{d\/\varrho/d\/\nu}\bigr]$ is a minimizing subalgebra of each minimizing $R$. 
Thus it is the least one of this sort. 
\end{pf}
\subsubsection{${\mathcal R}_M(\nu,\varrho)$ as a projective subalgebra}\label{unter.333}
Suppose $\varrho\ll \nu$ such that a least minimizing subalgebra exists. As has been remarked in line of the 
previous proof the algebra ${\mathcal R}_M(\nu,\varrho)$ then has to be a minimizing {\em projective} subalgebra. 
Application of  
{\sc Lemma \ref{aux0}} then yields that provided ${\mathcal R}_M(\nu,\varrho)$ exists the latter  
has to equal to 
\begin{subequation}\label{suppdef}  
\begin{equation}\label{suppdef.0}
R_\infty(\nu,\varrho)=\bigcap_{k\in 
s(\nu)^\perp M_+ s(\nu)^\perp} R[\sqrt{d\/\varrho/d\/\nu}+k]\,.
\end{equation}
From {\sc Lemma \ref{aux0}}\,(\ref{aux0.1a}) even ${\mathcal R}_M(\nu,\varrho)=
R[\sqrt{d\/\varrho/d\/\nu}+k_\infty]$ can be seen to hold, for some $k_\infty\in s(\nu)^\perp M_+ s(\nu)^\perp$.  
In line with (\ref{suppdef.0}) the latter especially means that  
$R[\sqrt{d\/\varrho/d\/\nu}+k_\infty]\subset R[\sqrt{d\/\varrho/d\/\nu}+
\lambda\,s(\nu)^\perp]$ has to be fulfilled, for each 
$\lambda\in {\mathbb R}_+$. Therefore $k_\infty\in {\mathbb R}_+\,s(\nu)^\perp$ has to hold. 
In summarizing from the latter and (\ref{suppdef.0}), in the general case of $\varrho\ll\nu$ the 
conclusion of {\sc Theorem \ref{radcomm}}\,(\ref{radcomm1}) and formula (\ref{radcomm.1})  
generalize to the following implication, which must be fulfilled for some 
$\gamma\in {\mathbb R}_+$\,: 
\begin{eqnarray}\label{suppdef0.1}
{\mathcal R}_M(\nu,\varrho)\mbox{ \em exists} & \Longrightarrow & {\mathcal R}_M(\nu,\varrho)=
\bigcap_{\lambda\in {\mathbb R}_+}R[\sqrt{d\/\varrho/d\/\nu}+\lambda\,s(\nu)^\perp]\nonumber \\ 
&  & 
\makebox{ }=R[\sqrt{d\/\varrho/d\/\nu}+\gamma\,s(\nu)^\perp]  \\
& & \makebox{ }=R_\infty(\nu,\varrho)\,.\nonumber
\end{eqnarray}
\end{subequation}
To summarize from this, for given $\{\nu,\varrho\}$ obeying $\varrho\ll\nu$ the algebra 
$R_\infty(\nu,\varrho)$  can be regarded to be the only candidate for ${\mathcal R}_M(\nu,\varrho)$.  
Thereby, the $\gamma$ within (\ref{suppdef0.1}) will be made more explicit later. 

Note that in the special case of $\varrho\ll\nu$ with $s(\varrho)\in M^\nu$ 
one can go a step further. Then, since owing to 
{\em Example \ref{zentral}} the assertion of 
{\sc Lemma \ref{aux0}}\,(\ref{aux0.1a}) can be applied to any minimizing 
subalgebra $R$, the above 
can be strengthened to the assertion that, depending from whether or not $R_\infty(\nu,\varrho)$ is 
minimizing, either a least minimizing abelian subalgebra will 
exist and then obeys ${\mathcal R}_M(\nu,\varrho)=R_\infty(\nu,\varrho)$, 
or a least minimizing abelian subalgebra cannot exist at all.
\begin{lemma}\label{suppdef.0a}
Suppose $\varrho\ll\nu$, with $s(\varrho)\in M^\nu$. Then $R_\infty(\nu,\varrho)$ 
is minimizing if, and only if, a least minimizing abelian subalgebra exists. 
\end{lemma} 
Having in mind these facts, and 
knowing that the special case of faithful $\nu$ has been dealt with yet 
in {\sc Theorem \ref{radcomm}}, with providing a complete answer for faithful $\varrho$,  
we are now going to analyze the family of algebras occuring under the intersection within 
(\ref{suppdef0.1}) more thoroughly 
in the remaining cases (in particular, those with non-faithful $\nu$) which are not yet covered by the 
premises of {\sc Theorem \ref{radcomm}}. 
To this sake some auxiliary technical facts on 
hereditary subalgebras and elementary spectral theory will be needed. Recall 
some standard fact from ${\mathsf W}^*$-theory first.  
\begin{remark}\label{wstern.1}
If $R[y,y^*]$ is the smallest ${\mathsf W}^*$-subalgebra 
of $M$ generated by $y\in M$ and ${\mathbf 1}$, then this is the $\sigma(M,M_*)$-closure of 
all polynomials in $y,\,y^*$ (including the constants as ${\mathbb C}\cdot{\mathbf 1}$). 
Here, $M_*$ is the {\em predual} of $M$, which is 
the Banach (sub)space of $M^*$ (with respect to the functional norm) which is 
generated by all {\em normal} positive linear forms (refer also to the elements of $M_*$ as {\em normal 
\textup{(}linear\textup{)} forms}). The $\sigma(M,M_*)$-topology is 
the weakest 
locally convex topology on $M$ such that all the seminorms $p_f$, $f\in M_*$, with $p_f(x)=|f(x)|$ for $x\in M$, 
are continuous. 
\end{remark}
Suppose now $\varrho\ll\nu$, and let an orthoprojection $q$ be defined by 
$q=s(\varrho)+s(\nu)^\perp$. On the hereditary ${\mathsf W}^*$-subalgebra $qMq$ 
define another normal positive linear forms $\nu_q,\varrho_q$ by $\nu_q=\nu|_{qMq}$ and 
$\varrho_q=\varrho|_{qMq}$, respectively. Then $\varrho_q\ll\nu_q$ is fulfilled, with supports in 
$qMq$ obeying $s(\nu_q)=s(\varrho_q)=s(\varrho)$ and $s(\nu_q)^\perp=s(\nu)^\perp$, with `$\perp$' 
referring to $qMq$ and $M$, accordingly.   
Also, if $x=\sqrt{d\/\varrho/d\/\nu},\,x_q=\sqrt{d\/\varrho_q/d\/\nu_q}$ are the corresponding 
Radon-Nikodym operators one has $x_q=x$ as elements 
of $M$. Also, if ${\mathop{spec}}_p(x)$ and ${\mathop{spec}}_p(x_q)$ are the 
point-spectra of $x$ and $x_q=x$ with respect to $M$ and 
$qMq$, respectively, then the relation 
\begin{subequation}\label{herM}
\begin{equation}\label{herM.1}
{\mathop{spec}}_p(x_q)\cup \{0\}={\mathop{spec}}_p(x)
\end{equation}
can be easily seen to hold. For $y\in (qMq)_{\mathrm h}\subset M_{\mathrm h}$ we let  
$R_q[y]$ be the ${\mathsf W}^*$-subalgebra of $qMq$ generated by $y$ and the unity 
$q$ of $qMq$. In view of {\em Remark \ref{wstern.1}} it is plain to see that   
$R_q[y]=qR[y]q$ holds. We are going to show that provided ${\mathcal R}_M(\nu,\varrho)$ exists then  
${\mathcal R}_{qMq}(\nu_q,\varrho_q)$ exists and obeys 
\begin{equation}\label{herM.2}
{\mathcal R}_{qMq}(\nu_q,\varrho_q)=q{\mathcal R}_M(\nu,\varrho)q\,.
\end{equation}
\end{subequation}
In fact, since owing to $s(x)=s(\varrho)$ for each $k\in s(\nu)^\perp Ms(\nu)^\perp$ also 
$x+k\in qMq$ 
holds, one has $R_q[x_q+k]=qR[x+k]q$. Hence, in accordance with (\ref{suppdef.0}) and 
(\ref{suppdef0.1}) one has 
$q{\mathcal R}_M(\nu,\varrho)q=\cap_{\lambda\geq 0}R_q[x_q+\lambda\,s(\nu)^\perp]=
R_q[x_q+\gamma\,s(\nu)^\perp]=\cap_{k}R_q[x_q+k]$, for some real $\gamma\geq 0$. We may apply 
formula (\ref{suppdef.0}) with respect to the hereditary algebra $qMq$ and normal 
positive linear forms $\nu_q,\varrho_q$. The result is $R_\infty(\nu_q,\varrho_q)=\cap_{k}R_q[x_q+k]$, 
with $k$ running through $s(\nu_q)^\perp M_+s(\nu_q)^\perp=s(\nu)^\perp M_+s(\nu)^\perp$ 
(see above). Hence, in view of the previous one has $q{\mathcal R}_M(\nu,\varrho)q=
R_q[x_q+\gamma\,s(\nu)^\perp]=R_\infty(\nu_q,\varrho_q)$. 
Especially, application of {\sc Lemma \ref{aux0}}\,(\ref{aux0.1}) for $\nu_q,\varrho_q$ on $qMq$ 
then shows that $R_\infty(\nu_q,\varrho_q)$ is minimizing. But then, since $s(\nu_q)=s(\varrho_q)$ and 
$\varrho_q\ll \nu_q$ hold, when considering {\sc Lemma \ref{suppdef.0a}}, {\em Remark \ref{projrem}} (\ref{proj.2}) 
and (\ref{suppdef0.1}) for $\nu_q,\varrho_q$ on $qMq$, 
one gets $R_\infty(\nu_q,\varrho_q)={\mathcal R}_{qMq}(\nu_q,\varrho_q)$. 
From this in view of the above (\ref{herM.2}) follows.

Close our preliminaries with the following auxiliary result which matters some 
elementary spectral theory. 
\begin{lemma}\label{spec}  
Suppose $x\in M_+$, $s(x)<{\mathbf 1}$, with point spectrum ${\mathop{spec}}_p(x)$.  
Depending from the latter, the following cases may occur for the 
commutative ${\mathsf W}^*$-subalgebra 
$R_0(x)=\bigcap_{\lambda\in {\mathbb R}_+}R[x+\lambda\,s(x)^\perp]$,  
where $\gamma$ can stand for any non-negative real\textup{:}
$$
R_0(x)\,\left\{ 
\begin{array}{ll}
=R[x] & \mbox{if ${\mathop{spec}}_p(x)\backslash \{0\}=\emptyset$,}\\
 & \\  
=R[x+\lambda_0\,s(x)^\perp] & \mbox{if ${\mathop{spec}}_p(x)\backslash \{0\}=\{\lambda_0\}$,}\\
 & \\
\not=R[x+\gamma\,s(x)^\perp] & \mbox{if $\#{\mathop{spec}}_p(x)\backslash 
\{0\}\geq 2$.}\\
\end{array}
\right.
$$
Especially, $R_0(x)=R[x]$ holds if, and only if, 
${\mathop{spec}}_p(x)\backslash \{0\}=\emptyset$ is fulfilled.   
\end{lemma}
\smallskip
\begin{pf}
Some preliminary results will be derived first.   
Let $\{E_x(t):t\in {\mathbb R}\}$ be the spectral resolution  
of $x$ within the projection lattice of $M$. Then, the eigenprojection of the positive element 
$x+\lambda\,s(x)^\perp$ to the spectral value $\lambda\in {\mathbb R}_+$ is 
given by  
\begin{varequation}{\ensuremath{\star}}
E_{x+\lambda\,s(x)^\perp}(\{\lambda\})=\biggl\{
\begin{array}{ll}
s(x)^\perp+E_x(\{\lambda\}) & \mbox{for $\lambda\in {\mathbb R}_+\backslash \{0\}$}\,,\\
 & \\
s(x)^\perp & \mbox{for $\lambda=0\,.$}
\end{array}
\biggr.
\end{varequation}
In fact, by assumption $E_x(\{0\})=s(x)^\perp$ holds, and thus the part of 
($\star$) relating to $\lambda=0$ is valid. Also, for 
$\lambda\in {\mathbb R}_+\backslash \{0\}$ it is  
clear from $E_x(\{0\})E_x(\{\lambda\})={\mathbf 0}$ and the above that 
$p=s(x)^\perp+E_x(\{\lambda\})$ is an orthoprojection in $M$, and 
which obeys $(x+\lambda\,s(x)^\perp)p=\lambda\,p$. Note in this context that 
$E_x(\{\lambda\})$ is non-vanishing iff $\lambda \in {\mathop{spec}}_p(x)$. 
Also, for an orthoprojection $q\geq p$ one has $(q-p)\,s(x)^\perp={\mathbf 0}$ and 
$(q-p)\,E_x(\{\lambda\})={\mathbf 0}$. Hence, assuming  
$(x+\lambda\,s(x)^\perp)\,q=\lambda\,q$ yields $x\,(q-p)=\lambda\,(q-p)$, 
which according to spectral theory necessarily implies 
$(q-p)\leq  E_x(\{\lambda\})$. In view of the above then $(q-p)={\mathbf 0}$. 
Thus, there is no larger than $p$ orthoprojection $q$ in $M$ with 
$(x+\lambda\,s(x)^\perp)\,q=\lambda\,q$, which means 
$p=E_{x+\lambda\,s(x)^\perp}(\{\lambda\})$. This is ($\star$).\\ 
Next, it is useful to take notice that the following alternatives exist\textup{:}
\begin{varequation}{\ensuremath{\star\star}}
R[x+\lambda\,s(x)^\perp]\,\biggl\{ 
\begin{array}{ll}
=R[x] & \mbox{if $\lambda\not\in {\mathop{spec}}_p(x)\backslash \{0\}$ or $\lambda=0$,}\\
 & \\  
\subsetneqq R[x] & \mbox{else}.\\
\end{array}
\biggr.
\end{varequation}
To see ($\star\star$), note first that obviously 
$R[x+\lambda\,s(x)^\perp]\subset R[x]$. Since for 
$\lambda\not\in {\mathop{spec}}_p(x)\backslash \{0\}$ one has 
$E_x(\{\lambda\})={\mathbf 0}$, from ($\star$)  
then $E_{x+\lambda\,s(x)^\perp}(\{\lambda\})=s(x)^\perp$ is seen, and thus  
both $x+\lambda\,s(x)^\perp$ and $s(x)^\perp$ have to belong to  
$R[x+\lambda\,s(x)^\perp]$, and thus $x$ does, too. In view of the above 
then $R[x+\lambda\,s(x)^\perp]= R[x]$, which   
for $\lambda=0$ is trivially valid, is seen to hold for  
$\lambda\not\in {\mathop{spec}}_p(x)\backslash \{0\}$.  
In case of 
$\lambda\in {\mathop{spec}}_p(x)\backslash \{0\}$, the element  
$x+\lambda\,s(x)^\perp$ has full support, and according 
to ($\star$) $s(x)^\perp$ is a {\em proper} subprojection of the  
eigen-orthoprojection $E_{x+\lambda\,s(x)^\perp}(\{\lambda\})$ to the 
spectral value $\lambda\in {\mathop{spec}}_p(x+\lambda\,s(x)^\perp)$. Since each spectral 
eigenprojection has to be a minimal orthoprojections of the generated commutative 
${\mathsf W}^*$-algebra 
$R[x+\lambda\,s(x)^\perp]$, from the previous    
$s(x)^\perp\not\in R[x+\lambda\,s(x)^\perp]$ has to be followed. Hence, in this 
case then $R[x+\lambda\,s(x)^\perp]\subsetneqq R[x]$, which 
completes the proof of ($\star\star$).     

After these preparations, we are going to prove the assertions 
of our results on $R_0(x)$. Note that the validity in case of ${\mathop{spec}}_p(x)\backslash \{0\}=\emptyset$ or 
${\mathop{spec}}_p(x)\backslash \{0\}=\{\lambda_0\}$ is straightforward  
from ($\star\star$). Thus, we have to consider  
explicitely only the case with     
$\#{\mathop{spec}}_p(x)\backslash \{0\}\geq 2$.  From ($\star\star$)  
then obviously $R_0(x)\subsetneqq R[x]$ follows. Especially this also means that 
the assertion is valid for $\gamma=0$. Now, in line with this, but in contrast with the assertion, 
assume we had $R_0(x)=R[x+\gamma\,s(x)^\perp]$, with $\gamma>0$. 
Then, since $\#{\mathop{spec}}_p(x)\backslash \{0\}\geq 2$ is fulfilled, there has to 
exist $\lambda\in {\mathop{spec}}_p(x)\backslash \{0\}$ with $\lambda\not=\gamma$. 
Thus $E_{x+\lambda\,s(x)^\perp}(\{\lambda\})\in R[x+\lambda\,s(x)^\perp]$, and 
$E_{x+\gamma\,s(x)^\perp}(\{\gamma\})\in R_0(x)$ by assumption. Since by definition of $R_0(x)$ 
one has   
$R_0(x)\subset R[x+\lambda\,s(x)^\perp]$, both $E_{x+\lambda\,s(x)^\perp}(\{\lambda\})$ and 
$E_{x+\gamma\,s(x)^\perp}(\{\gamma\})$ have to be in $R[x+\lambda\,s(x)^\perp]$.   
From ($\star$) and 
since $\gamma\not=\lambda$ is fulfilled, we see $s(x)^\perp=
E_{x+\gamma\,s(x)^\perp}(\{\gamma\})\,E_{x+\lambda\,s(x)^\perp}(\{\lambda\})\in 
R[x+\lambda\,s(x)^\perp]$, and therefore also 
$x\in R[x+\lambda\,s(x)^\perp]$ holds. From this and 
$R[x+\lambda\,s(x)^\perp]\subset R[x]$ then 
$R[x+\lambda\,s(x)^\perp]=R[x]$ had to be followed. Owing to 
the choice of $\lambda$ in accordance with $\lambda\in {\mathop{spec}}_p(x)\backslash \{0\}$ 
this is in contradiction with ($\star\star$). Thus, also in case of $\gamma>0$ a relation 
$R_0(x)=R[x+\gamma\,s(x)^\perp]$ cannot happen. 
Finally, note that by the just proven allowance is made 
for any situations with $R_0(x)$ that might occur. Particularly, from this and   
($\star\star$) one also infers that $R_0(x)=R[x]$ cannot happen unless  
${\mathop{spec}}_p(x)\backslash \{0\}=\emptyset$, whereas in the latter case 
this then in fact occurs. Thus, also the final assertion is seen to be true.             
\end{pf}
\subsubsection{The main result for $\varrho\ll\nu$ and with $s(\varrho)\in M^\nu$}\label{unter.334}
Suppose $\varrho\ll\nu$ such that ${\mathcal R}_M(\nu,\varrho)$ exists. Then, we derive a  
formula of ${\mathcal R}_M(\nu,\varrho)$ which generalizes (\ref{radcomm.1}) to this context. In addition 
also partial answers on the existence problem for ${\mathcal R}_M(\nu,\varrho)$ will be given. 
\begin{theorem}\label{suppdef.2}
Let $M$ be a ${\mathsf W}^*$-algebra, and let two normal positive linear forms $\nu,\varrho$ be given 
on $M$ and obeying $\varrho\ll\nu$.  
Let a non-negative real $\lambda_0$ be defined by 
\begin{subequation}\label{suppdef.20}
\begin{equation}\label{gpv}
\lambda_0=\sup\biggl\{\lambda:\,\lambda\in {\mathop{spec}}_p(\sqrt{d\/\varrho/d\/\nu})\cup\{0\}\biggr\}\,.
\end{equation} 
The following facts hold true. 
\begin{enumerate}[0000]
\item\label{suppdef.2.1}
Provided ${\mathcal R}_M(\nu,\varrho)$ exists then it obeys
\begin{equation}\label{radcomm.11}
{\mathcal R}_M(\nu,\varrho)=R\biggl[\sqrt{d\/\varrho/d\/\nu}+\lambda_0\,s(\nu)^\perp\biggr]\,,
\end{equation}
with the additional condition   
\begin{equation}\label{radokondi}
\# {\mathop{spec}}_p(\sqrt{d\/\varrho/d\/\nu})
\,\biggl\{ 
\begin{array}{ll}
\leq 2 & \mbox{if $s(\varrho)=s(\nu)$,}\\
 & \\  
= 1 & \mbox{else}\\
\end{array}
\biggr.
\end{equation} 
fulfilled in case of non-faithful $\nu$.
\item\label{suppdef.2.1a}
Assume $\{\nu,\varrho\}$ with $s(\varrho)\in M^\nu$. Then, if $\nu$ is faithful, or 
in all cases with 
non-faithful $\nu$ obeying $\dim s(\nu)^\perp M s(\nu)^\perp<\infty$ and $\varrho$ respecting  
\textup{(\ref{radokondi})}, a least minimizing abelian subalgebra exists.   
\end{enumerate}
\end{subequation}
\end{theorem}
\smallskip
\begin{pf}
Let $x=\sqrt{d\/\varrho/d\/\nu}$ and assume ${\mathcal R}_M(\nu,\varrho)$ exists. 
Then, (\ref{suppdef0.1}) yields that 
${\mathcal R}_M(\nu,\varrho)=
R[x+\gamma\,s(\nu)^\perp]$ has to be fulfilled, for  
some $\gamma\in {\mathbb R}_+$. We are going to determine the real $\gamma$ in terms of $x$. 
Let $q=s(\varrho)+s(\nu)^\perp$. According to (\ref{herM.2}) and in  
using the notations introduced in context of eqs.\,(\ref{herM}), with  
respect to the hereditary ${\mathsf W}^*$-subalgebra $qMq$ and normal positive linear forms 
$\nu_q,\varrho_q$ then 
also ${\mathcal R}_{qMq}(\nu_q,\varrho_q)$ exists and obeys ${\mathcal R}_{qMq}(\nu_q,\varrho_q)=
R_q[x_q+\gamma\,s(\nu)^\perp]$. On the other hand, an application of (\ref{suppdef0.1}) on 
$qMq$ with $\nu_q,\varrho_q$ yields ${\mathcal R}_{qMq}(\nu_q,\varrho_q)=R_0(x_q)$,  
with the algebra $R_0(x_q)$ constructed as in {\sc Lemma \ref{spec}} in terms of 
$x_q=\sqrt{d\/\varrho_q/d\/\nu_q}$ and with respect to $qMq$. Since both $s(x_q)=s(\varrho_q)=s(\nu_q)
=s(\varrho)$ and $s(\nu_q)^\perp=s(\nu)^\perp$ hold on $qMq$, 
in view of the above we therefore conclude that, 
provided ${\mathcal R}_M(\nu,\varrho)$ has been assumed to exist, then 
$R_0(x_q)=R_q[x_q+\gamma\,s(x_q)^\perp]$ has to be fulfilled, for some 
$\gamma\in {\mathbb R}_+$. But then, in case of $s(\varrho)=s(x_q)<q$, {\sc Lemma \ref{spec}} 
can be applied on $qMq$ and gives that $\# {\mathop{spec}}_p(x_q)\backslash\{0\}< 2$ has to be 
fulfilled, with $\gamma=\sup\{\lambda:\,\lambda\in {\mathop{spec}}_p(x_q)\cup\{0\}\}$. 
Note that the 
condition $s(\varrho)=s(x_q)<q$ is equivalent with $s(\nu)<{\mathbf 1}$, and that in this case 
then $0\in {\mathop{spec}}_p(x)$ holds. Hence, by (\ref{herM.1}) in this case 
$\# {\mathop{spec}}_p(x_q)\backslash\{0\}=\# {\mathop{spec}}_p(x)\backslash\{0\}$. Especially, 
the previously given $\gamma$ then obeys $\gamma=\lambda_0$, with $\lambda_0$ as 
given in accordance with (\ref{gpv}). Thus, 
in summarizing from this and the previous,   
assuming that ${\mathcal R}_M(\nu,\varrho)$ exists for non-faithful $\nu$ implies that    
(\ref{radcomm.11}) and $\# {\mathop{spec}}_p(x)\leq 2$ hold.  
Now, suppose $s(\varrho)<s(\nu)<{\mathbf 1}$. Then, assuming $\lambda_0>0$ would imply 
$q^\perp\in R[x+\lambda_0\,s(\nu)^\perp]$, for $q^\perp$ is the eigenprojection of 
$x+\lambda_0\,s(\nu)^\perp$ to eigenvalue $0$. But at the same time certainly $q^\perp\not\in 
R[x]$ since by supposition of this case $q^\perp<s(\varrho)^\perp$ has to hold and 
$s(\varrho)^\perp$ has to be a minimal orthoprojection of $R[x]$. Thus,  
$R[x+\lambda_0\,s(\nu)^\perp]$ cannot be a subalgebra of $R[x]$ in this case. In view of 
the meaning of ${\mathcal R}_M(\nu,\varrho)$ and since $R[x]$ is minimizing 
the latter contradicts the just derived formula (\ref{radcomm.11}) in the case of non-faithful $\nu$.
Hence, for $s(\varrho)<s(\nu)<{\mathbf 1}$ one must have $\lambda_0=0$. In view of 
(\ref{gpv}) and since for non-faithful $\nu$ one has $0\in {\mathop{spec}}_p(x)$ one then infers that 
${\mathop{spec}}_p(x)=\{0\}$ holds. This completes the proof of (\ref{radokondi}).     
That (\ref{radcomm.11}) remains true also 
for faithful $\nu$ follows 
since then owing to $s(\nu)^\perp={\mathbf 0}$ formula (\ref{radcomm.11}) simply reduces to 
formula (\ref{radcomm.1}), which according to {\sc Theorem \ref{radcomm}}\,(\ref{radcomm1}) is true, 
however, and which completes the proof of (\ref{suppdef.2.1}).

To see (\ref{suppdef.2.1a}), note that for faithful $\nu$ formula (\ref{suppdef.0}) yields 
$R_\infty(\nu,\varrho)=R[\sqrt{d\/\varrho/d\/\nu}]$. Hence, according to {\sc Lemma \ref{aux0}},  
the algebra $R_\infty(\nu,\varrho)$ is minimizing. 
But then, since $\varrho$ obeys $s(\varrho)\in M^\nu$, from {\sc Lemma \ref{suppdef.0a}} we may  
also conclude that ${\mathcal R}_M(\nu,\varrho)$ exists. This proves the part of 
(\ref{suppdef.2.1a}) relating to a faithful $\nu$.\\ 
Suppose now $\nu$ to be non-faithful, but with $\dim s(\nu)^\perp M s(\nu)^\perp<\infty$ fulfilled, and 
$\varrho$ such that $s(\varrho)\in M^\nu$ holds and condition (\ref{radokondi}) is respected. 
Note that in this 
case $0\in {\mathop{spec}}_p(x)$ holds. Also, by the assumption of finite dimensionality then 
${\mathop{spec}}(k)={\mathop{spec}}_p(k)$ holds, for each $k\in s(\nu)^\perp M_+ s(\nu)^\perp$, 
and if $p_\lambda$ is the eigenprojection of $k$ to 
$\lambda\in {\mathop{spec}}_p(k)$, we have 
$
\sum_{\lambda\in {\mathop{spec}}_p(k)} p_\lambda=
s(\nu)^\perp$.  
By the same kind of auxiliary arguments from 
elementary spectral theory, which have been yet using in line of the proof 
of {\sc Lemma \ref{spec}} in some special case,     
in literally the same way (the details of which therefore will not be mentioned) 
can be also applied in order to compare the spectral structures of $x+k$ and $x$ 
(below these facts will be tacitly made use of). Suppose $\lambda_0=0$ first. Then, zero is the only 
eigenvalue of $x$, and therefore one infers that  
${\mathop{spec}}_p(x+k)={\mathop{spec}}_p(k)\cup \{0\}$ for $s(\varrho)<s(\nu)$, and 
${\mathop{spec}}_p(x+k)={\mathop{spec}}_p(k)$ for $s(\varrho)=s(\nu)$. Owing to this and 
$s(x)\leq s(\nu)$, whereas each of the above $p_\lambda$ for 
$\lambda \in {\mathop{spec}}_p(k)\backslash \{0\}$ 
will be also the corresponding 
eigenprojection to the same $\lambda
\in{\mathop{spec}}_p(x+k)$ with respect to $x+k$, the projection $p_0+\{s(\nu)-s(\varrho)\}$, or 
$\{s(\nu)-s(\varrho)\}$ respectively, will be the 
eigenprojection of $x+k$ to eigenvalue zero in case of $0\in {\mathop{spec}}_p(x+k)\cap
{\mathop{spec}}_p(k)$, and in case of $0\in {\mathop{spec}}_p(x+k)$ but with   
$0\not\in {\mathop{spec}}_p(k)$, respectively. Therefore, $p_\lambda\in R[x+k]$ for each 
$\lambda\in {\mathop{spec}}_p(k)\backslash\{0\}$, and $p_0+\{s(\nu)-s(\varrho)\}\in R[x+k]$ 
in case of $0\in {\mathop{spec}}_p(x+k)\cap{\mathop{spec}}_p(k)$ or 
$\{s(\nu)-s(\varrho)\}\in R[x+k]$ in case of $0\in {\mathop{spec}}_p(x+k)$ but with   
$0\not\in {\mathop{spec}}_p(k)$. But then in view of the above in each case also their sum 
$s(\nu)^\perp+\{s(\nu)-s(\varrho)\}$ has to be in $R[x+k]$, that is, $s(\varrho)^\perp\in R[x+k]$ 
has to hold. From this owing to $s(x)=s(\varrho)\leq s(\nu)$ then $x=s(\varrho)\{x+k\}\in R[x+k]$ is seen. 
Hence, $R[x+k]\supset R[x]$ follows, for each $k\in s(\nu)^\perp M_+ s(\nu)^\perp$, and therefore 
one has $R_\infty(\nu,\varrho)=R[x]$. From {\sc Lemma \ref{aux0}} follows that   
$R_\infty(\nu,\varrho)$ is minimizing.  Thus, since $\varrho$ obeys 
$s(\varrho)\in M^\nu$, by {\sc Lemma \ref{suppdef.0a}} we may 
conclude that ${\mathcal R}_M(\nu,\varrho)$ exists. Hence,   
for non-faithful $\nu$ and $\# {\mathop{spec}}_p(x)=1$ the assertion of (\ref{suppdef.2.1a}) is true. 

Suppose $s(\varrho)=s(\nu)$ and $\# {\mathop{spec}}_p(x)=2$, with non-faithful $\nu$. 
Then, $\lambda_0>0$, and for each 
$k\in s(\nu)^\perp M_+ s(\nu)^\perp$ one has 
$p_\lambda\in R[x+k]$, for $\lambda\in {\mathop{spec}}_p(k)\backslash\{\lambda_0\}$. If $\lambda_0\not\in 
{\mathop{spec}}_p(k)$ from this $s(\nu)^\perp\in R[x+k]$ follows, from which $R[x]\subset R[x+k]$ 
is seen. For $\lambda_0\in {\mathop{spec}}_p(k)$, however, $p_{\lambda_0}+E_x(\{\lambda_0\})$ is 
the $\lambda_0$ corresponding eigenprojection of $x+k$, and therefore instead of 
$p_{\lambda_0}\in R[x+k]$ 
one finds $p_{\lambda_0}+E_x(\{\lambda_0\})\in R[x+k]$. Summing up then yields 
$s(\nu)^\perp+E_x(\{\lambda_0\})\in R[x+k]$ instead. But then also 
$k+\lambda_0\,E_x(\{\lambda_0\})=(s(\nu)^\perp+E_x(\{\lambda_0\}))(x+k)\in 
R[x+k]$. Hence, since $x+\lambda_0\,s(\nu)^\perp$ can be combined together from 
the mentioned elements as  
$x+\lambda_0\,s(\nu)^\perp=(x+k)-(k+\lambda_0\,E_x(\{\lambda_0\}))+
\lambda_0(s(\nu)^\perp+E_x(\{\lambda_0\}))$, 
$x+\lambda_0\,s(\nu)^\perp\in R[x+k]$ is seen. 
Note that owing to $s(\nu)^\perp\in R[x]$ in any case one has $x+\lambda_0\,s(\nu)^\perp\in R[x]$. 
We may summarize these facts and conclude that, for non-faithful $\nu$ with  
$s(\varrho)=s(\nu)$ and $\# {\mathop{spec}}_p(x)=2$, $R[x+\lambda_0\,s(\nu)^\perp]\subset R[x+k]$ 
holds, for each $k\in s(\nu)^\perp M_+ s(\nu)^\perp$. Hence, 
$R_\infty(\nu,\varrho)=R[x+\lambda_0\,s(\nu)^\perp]$, and thus 
according to {\sc Lemma \ref{aux0}} also in this case   
the algebra $R_\infty(\nu,\varrho)$ is minimizing. Since  
$s(\varrho)=s(\nu)\in M^\nu$ holds {\sc Lemma \ref{suppdef.0a}} can be applied once more again and 
yields that ${\mathcal R}_M(\nu,\varrho)$ exists. This closes the proof of  
(\ref{suppdef.2.1a}), and at the same time also completes the 
proof of the theorem.
\end{pf}
\subsubsection{Examples and consequences}\label{unter.335} 
Start with discussing {\sc Theorem \ref{suppdef.2}} in the finite dimensional case. 
\begin{exam}\label{ab3ex}
Suppose $2\leq\dim M<\infty$, and $\nu,\varrho$ two non-zero positive linear forms obeying $\varrho\ll \nu$, 
but which are not mutually proportional. Then, 
the corresponding Radon-Nikodym operator cannot be proportional 
to the support of $\nu$, $\sqrt{d\/\varrho/d\/\nu}\not\in {\mathbb R}_+\,s(\nu)$. 
Since $s(\varrho)\leq s(\nu)$ is the support of $\sqrt{d\/\varrho/d\/\nu}$, from these facts 
$\#\mathop{spec}(\sqrt{d\/\varrho/d\/\nu})\geq 2$ follows. Hence, 
since by finite-dimensionality one has 
$\mathop{spec}_p(\sqrt{d\/\varrho/d\/\nu})=\mathop{spec}(\sqrt{d\/\varrho/d\/\nu})$, the 
condition (\ref{radokondi}) in case of non-faithful $\nu$ could be satisfied only if 
$\#\mathop{spec}(\sqrt{d\/\varrho/d\/\nu})=2$ and $s(\varrho)=s(\nu)<{\mathbf 1}$ were 
fulfilled. But then $\sqrt{d\/\varrho/d\/\nu}$ as a Radon-Nikodym operator 
had to be proportional with $s(\varrho)=s(\nu)$, which however contradicts to the above mentioned fact. 
Thus, in view of {\sc Theorem \ref{suppdef.2}}\,(\ref{suppdef.2.1}) for 
non-faithful $\nu$ and under the above premises 
a least minimizing algebra cannot exist in the finite dimensional case.  
Especially, from the latter and by formula (\ref{radcomm.11}) one also infers 
that provided a least minimizing algebra exists then 
${\mathcal R}_M(\nu,\varrho)=R[\sqrt{d\/\varrho/d\/\nu}]$ will occur, in any case. 
From {\sc Theorem \ref{suppdef.2}}\,(\ref{suppdef.2.1a}) one infers that 
the latter case really can happen, e.g. in case of faithful 
$\nu$ and $\varrho$ obeying $\varrho\ll \nu$ and $s(\varrho)\in M^\nu$. 
\end{exam}
As the previous example shows the deviation from the law (\ref{radcomm.1}) as indicated by (\ref{radcomm.11}) 
could be observed only for $\dim M=\infty$. 
That this deviation really can occur is seen by the following example. 
\begin{exam}\label{endegut}
Let $M={\mathsf L}^\infty(I,m^\prime)$, where $\{I,m^\prime\}$ is the unit interval $I=[0,1]$   
with a measure $m^\prime=(m+\delta_0)/2$, where $m$ is the Lebesgue measure and $\delta_0$ 
is concentrated on $\{0\}$, with $\delta_0(\{0\})=1$. Let $\nu$ correspond to 
the class of the 
characteristic function $\chi_{(0,1]}$ of $(0,1]$ via $\nu(\cdot)=\int_{(0,1]} (\cdot)\, d\/m^\prime$, and be 
$f$ a strictly increasing function, which is continuous on $[0,1]$, except for one point 
$t_0>0$ where it is only left-continuous with $f(t_0)=\lambda_0>0$, and which obeys $0<f(t)\leq 1$ 
for $t>0$, and $f(0)=0$. Define $\varrho(\cdot)=\int_I (\cdot)f\,d\/m^\prime$. 
Then, $\varrho\ll \nu$ (even $\varrho\leq \nu$ holds) and 
$s(\nu)=s(\varrho)=\chi_{(0,1]}<\chi_{[0,1]}={\mathbf 1}$, with Radon-Nikodym operator 
$x=f$ obeying $\{0,\lambda_0\}=
\mathop{spec}_p(x)$. Hence condition (\ref{radokondi}) is fulfilled in this case. 
Since owing to $M^\nu=M$ one has $s(\varrho)\in M^\nu$ to be fulfilled by triviality, 
{\sc Theorem \ref{suppdef.2}}\,(\ref{suppdef.2.1a}) can be applied 
and formula (\ref{radcomm.11}) then yields ${\mathcal R}_M(\nu,\varrho)=R[f+\lambda_0\,\chi_{\{0\}}]$.  
\end{exam}
Along with {\sc Theorem \ref{suppdef.2}}\,(\ref{suppdef.2.1}) comes   
another necessary condition for ${\mathcal R}_M(\nu,\varrho)$ to exist 
which often will be useful. To explain this, in 
the following let ${\mathsf{Aut}}(M)$ denote the group of all 
$^*$-automorphisms of $M$, and for $y\in M$ we let  
${\mathsf{Aut}}_y(M)$ be those $^*$-automorphisms which leave the element 
$y$ fixed. Clearly, since we have to do with $^*$-automorphisms one has 
${\mathsf{Aut}}_y(M)={\mathsf{Aut}}_{y^*}(M)$, for each $y\in M$.  
\begin{remark}\label{wstern.2}
Remind that a $^*$-isomorphism $\Phi$ from one ${\mathsf W}^*$-algebra $M$ onto another 
${\mathsf W}^*$-algebra $N$ automatically is $\sigma(M,M_*)$-$\sigma(N,N_*)$ continuous. 
From this and {\em Remark \ref{wstern.1}} then follows that  
$\Phi\in {\mathsf{Aut}}_y(M)\ \Longleftrightarrow\ \Phi\in {\mathsf{Aut}}_x(M),\,
\forall\,x\in R[y,y^*]$,  
is valid for each $y\in M$.  
\end{remark}
\begin{corolla}\label{symme}   
For the pair $\{\nu,\varrho\}$ of normal positive linear forms suppose $\varrho\ll\nu$, 
with Radon-Nikodym operator $x=\sqrt{d\/\varrho/d\/\nu}$,   
and let $\lambda_0$ be defined in accordance with formula  
\textup{(\ref{gpv})}. Then, existence of ${\mathcal R}_M(\nu,\varrho)$ 
implies the following to hold\textup{:}     
\begin{equation}\label{symme.1}
\forall\,k\in s(\nu)^\perp M_+s(\nu)^\perp:\ 
{\mathsf{Aut}}_{x+k}(M)\subset{\mathsf{Aut}}_{x+\lambda_0\,s(\nu)^\perp}(M)\,.
\end{equation}
\end{corolla}
\smallskip
\begin{pf}       
In view of (\ref{suppdef.0}) and 
{\sc Theorem \ref{suppdef.2}}\,(\ref{suppdef.2.1})  
the premises imply    
$R[x+\lambda_0\,s(\nu)^\perp]\subset R[x+k]$ to be fulfilled, for each 
$k\in s(\nu)^\perp M_+s(\nu)^\perp$. From this it is evident that by each  
$^*$-automorphisms $\Phi$ leaving pointwise invariant all elements of 
$R[x+k]$ in particular also each element of 
$R[x+\lambda_0\,s(\nu)^\perp]$ is left invariant. This is (\ref{symme.1}).  
\end{pf}
We will show that  
among the assumptions in {\sc Theorem \ref{suppdef.2}}\,(\ref{suppdef.2.1a}) 
also the condition 
$\dim s(\nu)^\perp M s(\nu)^\perp<\infty$ is a sensitive one. For simplicity 
this will be demonstrated by such an example, which by its construction and owing to the procedure 
applied can stand 
for a whole class of analogous (even non-commutative) situations where 
(\ref{symme.1}) fails and thus a least minimizing subalgebra cannot exist then.
\begin{exam}\label{symme.2}        
Let $M={\mathsf L}^\infty(I,m)$, where $\{I,m\}$ is the unit interval $I=[0,1]$   
with Lebesgue-measure $m$. Let $\tau\in M_+^*$ be the standard tracial state 
given on $M$ by $\tau(x)=\int_I d\/m\,x$, for $x\in M$. Suppose $\nu=\tau(\chi_0(\cdot))$, 
where $\chi_0$ corresponds to the class of the characteristic function of the interval 
$[0,1/2]$. Assume $\varrho=\tau(f(\cdot))$, where we let $f$ correspond to the class of 
some continuous, monotoneous function $f$ on $[0,1]$, with $1\geq f(t)>0$ for $t<1/2$ and $f(t)=0$ 
else. We then have $\varrho\ll \nu$, 
$s(\nu)=\chi_0<{\mathbf 1}$ and  
$x=\sqrt{d\/\varrho/d\/\nu}=f$. Let us consider the $^*$-automorphism 
$\Phi_g$ which is induced on $M$ by the measure preserving 
point-transformation $g:\,I\ni t\,\mapsto (1-t)\in I$ of the unit interval, 
that is, in the sense of equivalence of functions, 
$\Phi_g(x)=x\circ g$ is fulfilled. Obviously, $\Phi_g$ is idempotent, that is, 
a symmetry. Note that $\Phi_g(\chi_0)=\chi_1$ holds, 
where $\chi_1$ stands for the class of the characteristic function of the 
interval $[1/2,1]$ within $M$, that is, $\Phi_g(\chi_0)=\chi_0^\perp$ is 
fulfilled. From $0\leq f\leq \chi_0$ 
then $\Phi_g(f)\in\chi_0^\perp M_+ \chi_0^\perp$ follows. Let us define 
$k=\Phi_g(f)$. Owing to idempotency of $\Phi_g$ then   
$\Phi_g\in {\mathsf{Aut}}_{x+k}(M)$ follows. On the other 
hand, according to the above and since $\chi_0\in R[x]$ holds we certainly 
must have $\Phi_g\not\in {\mathsf{Aut}}_x(M)$. In fact, otherwise according to the equivalence   
mentioned on in {\em Remark \ref{wstern.2}}, in contrast to the above we also had 
$\chi_0$ to be a fixed point of $\Phi_g$, a contradiction.  
Now, the Radon-Nikodym operator $x=f$ by choice of $f$ obeys 
${\mathop{spec}}_p(x)=\{0\}$. Hence, $\lambda_0=0$. But then existence of 
the above constructed $\Phi_g$ proves that condition 
(\ref{symme.1}) is violated, and thus in view of {\sc Corollary \ref{symme}} this means 
that ${\mathcal R}_M(\nu,\varrho)$ cannot exist in the case at hand.  
\end{exam}
\subsubsection{Does each minimizing subalgebra dominate a 
minimizing projective subalgebra\,?}\label{unter.336} 
Note that according to {\sc Theorem \ref{suppdef.2}\,(\ref{suppdef.2.1})} and 
{\sc Lemma \ref{aux0}\,(\ref{aux0.1})} existence of the least minimizing subalgebra especially also means  
that each minimizing subalgebra $R$ possesses a 
minimizing projective subalgebra. One finds the following useful 
auxiliary characterization of this fact. 
\begin{corolla}\label{opt.1}
Let $\nu,\varrho$ be normal positive linear forms, with $\varrho\ll\nu$ and Radon-Nikodym operator $x$. 
Let $R$ be a minimizing abelian ${\mathsf W}^*$-subalgebra, and be $z\in R_+$ 
the $R$-relative Radon-Nikodym operator achieving $\varrho|_R=\nu|_R^z$.   
The following items are mutually equivalent\textup{:}
\begin{enumerate}[0000]
\item\label{opt.11}
\ $R_1\subset R$, for some minimizing projective subalgebra $R_1$\,;
\item\label{opt.12}
\ $\nu^{s(\varrho)}(z)=\nu(z)$\,.
\end{enumerate}
In the latter case $R_1=R[x+k]$ can be chosen in \textup{(\ref{opt.11})}, for some 
$k\in s(\nu)^\perp M_+ s(\nu)^\perp$.
\end{corolla}
\smallskip
\begin{pf}
For a minimizing $R$ the condition $\nu^{s(\varrho)}(z)=\nu(z)$ 
implies existence of $k\in s(\nu)^\perp M_+ s(\nu)^\perp$ 
with $R[x+k]\subset R$. This can be seen exactly along the same way as 
demonstrated in 
the course of the proof of 
{\sc Lemma \ref{aux0}\,(\ref{aux0.1a})} (see from (\ref{rel}) onward). 
In view of {\sc Lemma \ref{aux0}\,(\ref{aux0.1})} then $R_1=R[x+k]$ can be 
chosen in (\ref{opt.11}). 
To see the other direction, assume 
$R_1\subset R$ with some minimizing projective subalgebra $R_1$. 
By {\sc Lemma \ref{aux0}\,(\ref{aux0.1a})} one knows that $k\in s(\nu)^\perp M_+ s(\nu)^\perp$ 
exists with $R[x+k]\subset R_1$. Then also $R[x+k]\subset R$ holds, and thus   
$x+k\in R$. Owing to  
$s(\varrho|_R)\in R$ and since $R$ is commutative, one has 
$y=s(\varrho|_R)(x+k)=
(x+k)s(\varrho|_R)\in R_+$. From this and $\varrho=\nu^x=\nu^{(x+k)}$ then 
$\varrho|_R=\nu^{(x+k)}|_R=\nu^{(x+k)s(\varrho|_R)}|_R=\nu^y|_R=\nu|_R^y$ is seen.  
In view of $s(y)\leq s(\varrho|_R)$ and by uniqueness of the 
Radon-Nikodym operator $z$ in $R$ then $z=y$ follows.  
Now, $s(\varrho|_R)\geq s(\varrho)$ and $s(x)=s(\varrho)\leq s(\nu)$ 
hold.  
Hence, $s(\varrho)z=s(\varrho)y=s(\varrho)s(\varrho|_R)(x+k)=s(\varrho)(x+k)=x$ must be fulfilled, 
and therefore also $s(\varrho)z=zs(\varrho)=s(\varrho)zs(\varrho)$. Since $R$ is 
minimizing, from the previous together with {\sc Lemma \ref{opt}} (put $p=s(\varrho)$ and $a=z$ in 
(\ref{proj.-1})) by literally the same arguments which led us to 
see (\ref{rel}) within the proof of  
{\sc Lemma \ref{aux0}\,(\ref{aux0.1a})} then the   
desired relation $\nu^{s(\varrho)}(z)=\nu(z)$ is seen to hold also in the situation 
at hand.    
\end{pf}
\begin{remark}\label{opt.3}
\begin{enumerate}
\item\label{opt.31}
The condition $s(\varrho)\in M^\nu$ within {\sc Theorem \ref{suppdef.2}\,(\ref{suppdef.2.1a})} makes 
that {\sc Corollary \ref{opt.1}\,(\ref{opt.12})} is trivially satisfied, and then in line with   
{\em Remark \ref{projrem}}\,(\ref{proj.2}) each minimizing subalgebra is projective.
\item\label{opt.32}
Suppose $\varrho\ll \nu$ but with $s(\varrho)\not\in M^\nu$ (thus $M$ cannot be commutative). 
It is an open question whether other minimizing subalgebras than those respecting  
{\sc Corollary \ref{opt.1}\,(\ref{opt.12})} could exist at all. 
\item\label{opt.33}
Note that $M={\mathsf M}_2({\mathbb C})$ is the least case  
where the previous question might be non-trivial (cf.~{\em Example \ref{ab3ex}}). 
But in this case, the characteristic configuration of a pair $\{\nu,\varrho\}$ to be dealt with for a decision 
in the usual canonical 
manner may be reduced to pairs $\{a,p\}$ of $2\times 2$-matrices, with positive definite $a$ 
and one-dimensional orthoprojection $p$ obeying $pa\not=ap$. 
Thus calculations can be 
carried out explicitely (we omit 
the details), and in fact show that $R=R[x]=R[p]$ is the only minimizing subalgebra.   
This also completes the analysis of  
{\em Example \ref{ab3ex}}\, in the $2\times 2$-case: for  
$\nu,\varrho$ which are not mutually proportional and which obey 
$\varrho\ll\nu$ the least minimizing subalgebra 
exists iff $\nu$ is faithful. In view of 
{\em Example \ref{neg.2}}\, then even 
follows that, for a general pair of mutually non-proportional 
positive linear forms on 
$M={\mathsf M}_2({\mathbb C})$, ${\mathcal R}_M(\nu,\varrho)$ exists if, and only 
if, one of the two forms is faithful at least. Thus, in this case we have a complete 
solution of the problem for a non-commutative $M$, even without  
imposing the condition $\varrho\ll \nu$.      
\item\label{opt.34}
Suppose $\{\nu,\varrho\}$ such that {\sc Corollary \ref{opt.1}\,(\ref{opt.12})} be fulfilled in each case 
of a minimizing subalgebra. Then,    
the problem on existence of a least minimizing subalgebra will be reduced to 
the question whether or not 
$R_\infty(\nu,\varrho)$ were equal to $R[x+\lambda_0\,s(\nu)^\perp]$   
(see {\sc Lemma \ref{suppdef.0a}} for a special case). 
As {\em Example \ref{symme.2}}\, shows, for the latter 
to happen both (\ref{radokondi}) and (\ref{symme.1}) are 
necessary conditions, which are rather independent from each other. 
\item\label{opt.35}
The method by means of which the assertion on equality of the intersection algebra 
$R_\infty(\nu,\varrho)$ of (\ref{suppdef.0}) to one of the intersecting minimizing subalgebras 
$R[x+\lambda_0\,s(\nu)^\perp]$ has been disproved, and which is based on considering symmetries, 
seems to be very effective and in a modified form is a common method to disprove uniqueness of 
optimizing elements (algebras, decompositions,\,\ldots) in similar $^*$-algebraic optimization problems, 
see e.g.~\cite{Uhlm:98}.      
\end{enumerate}
\end{remark}
\theendnotes
\endnotetext{
Most of the material of section 2 as well as some parts of section 3, especially \ref{unter.32}, are reproduced from the part `foundational material' of the manuscript \cite{Albe:98}.}

\bibliographystyle{abbrv}

\begin{thebibliography}{10}

\bibitem{Albe:83}
P.~M. Alberti.
\newblock A note on the transition probability over ${C}^*$-algebras.
\newblock {\em Lett. Math. Phys.}, 7:25--32, 1983.

\bibitem{Albe:92.1}
P.~M. Alberti.
\newblock A {S}tudy on the {G}eometry of {P}airs of {P}ositive {L}inear
  {F}orms, {A}lgebraic {T}ransition {P}robability and {G}eometrical {P}hase
  over {N}on-{C}ommutative {O}perator {A}lgebras ({I}).
\newblock {\em Z. Anal. Anw.}, 11(3):293--334, 1992.

\bibitem{Albe:98}
P.~M. Alberti.
\newblock {Bures Geometry in the State Space of von-Neumann Algebras}.
\newblock (unpublished manuscript), 1998.

\bibitem{AlHe:89}
P.~M. Alberti and V.~Heinemann.
\newblock Bounds for the ${C}^*$-algebraic transition probability yield best
  lower and upper bounds to the overlap.
\newblock {\em J. Math. Phys.}, 30(9):2083--2089, September 1989.

\bibitem{AlUh:84}
P.~M. Alberti and A.~Uhlmann.
\newblock {Transition probabilities on $C^*$- and $W^*$-algebras}.
\newblock In H.~Baumg{\"a}rtel, G.~La{\ss}ner, A.~Pietsch, and A.~Uhlmann,
  editors, {\em {Proceedings of the Second International Conference on Operator
  Algebras, Ideals, and Their Applications in Theoretical Physics, Leipzig
  1983}}, pages 5--11, Leipzig, 1984. BSB B.G.Teubner Verlagsgesellschaft.
\newblock Teubner-Texte zur Mathematik, Bd.67.

\bibitem{Arak:71}
H.~Araki.
\newblock A remark on {Bures Distance Function} for normal states.
\newblock {\em Publ. RIMS (Kyoto)}, 6:477--482, 1970/71.

\bibitem{Arak:72}
H.~Araki.
\newblock Bures distance function and a generalization of {S}akai's
  non-commutative {R}adon-{N}ikodym theorem.
\newblock {\em Publ. RIMS (Kyoto)}, 8:335--362, 1972.

\bibitem{ArRa:82}
H.~Araki and G.~Raggio.
\newblock A remark on transition probability.
\newblock {\em Lett. Math. Phys.}, 6:237--240, 1982.

\bibitem{BeSh:55}
J.~Bendat and S.~Sherman.
\newblock Monotone and convex operator functions.
\newblock {\em Trans. Amer. Math. Soc.}, 79:58--71, 1955.

\bibitem{Buch:90}
D.~Buchholz.
\newblock On quantum fields that generate local algebras.
\newblock {\em J. Math. Phys.}, 31(8):1839--1846, August 1990.

\bibitem{Bure:69}
D.~Bures.
\newblock An extension of {K}akutani's theorem on infinite product measures to
  the tensor product of semifinite ${W}^*$ -algebras.
\newblock {\em Trans. Amer. Math. Soc.}, 135:199--212, 1969.

\bibitem{Cant:75}
V.~Cantoni.
\newblock Generalized transition probability.
\newblock {\em Comm. Math. Phys.}, 44(2):125--128, 1975.

\bibitem{Dixm:64}
P.~Dixmier.
\newblock {\em {Les $C^*$-Alg\`ebres et leurs Repr\'esentations}}.
\newblock Gauthier-Villars, Paris, 1964.

\bibitem{Dono:74}
W.~Donogue.
\newblock {\em Monotone matrix functions and analytic continuation}, volume 207
  of {\em Die Grundlehren der mathematischen Wissenschaften}.
\newblock Springer-Verlag, Berlin-Heidelberg-NY, 1974.

\bibitem{DyRu:66}
H.~Dye and B.~Russo.
\newblock A note on unitary operators in ${C}^*$-algebras.
\newblock {\em Duke Math. J.}, 33:413--416, 1966.

\bibitem{Gudd:81}
S.~Gudder.
\newblock Expectation and transition probability.
\newblock {\em Intern. J. Theor. Phys.}, 20:383--395, 1981.

\bibitem{Gudd:78}
S.~P. Gudder.
\newblock Cantoni's {Generalized Transition Probability}.
\newblock {\em Comm. Math. Phys.}, 63:265--267, 1978.

\bibitem{Haag:92}
R.~Haag.
\newblock {\em Local Quantum Physics}.
\newblock Texts and Monographs in Physics. Springer-Verlag,
  NY-Heidelberg-Berlin, 1992.

\bibitem{KaRi:83}
R.~Kadison and J.~Ringrose.
\newblock {\em Fundamentals of the Theory of Operator Algebras}, volume I
  Elementary Theory.
\newblock Academic Press, Inc., NY--London--Paris, 1983.

\bibitem{Mack:63}
G.~Mackey.
\newblock {\em The Mathematical Foundation of Quantum Mechanics}.
\newblock The Mathematical Physics Monograph Series. W.A.Benjamin, Inc., New
  York--Amsterdam, 1963.

\bibitem{Miel:69}
B.~Mielnik.
\newblock Theory of filters.
\newblock {\em Comm. Math. Phys.}, 15(1):1--46, 1969.

\bibitem{Saka:65}
S.~Sakai.
\newblock A {R}adon-{N}ikodym theorem in ${W}^*$-algebras.
\newblock {\em Bull. Amer. Math. Soc}, 71:149--151, 1965.

\bibitem{Saka:71}
S.~Sakai.
\newblock {\em ${C}^*$-Algebras and ${W}^*$-Algebras}.
\newblock Springer-Verlag, Berlin-Heidelberg-New York, 1971.

\bibitem{Uhlm:74}
A.~Uhlmann.
\newblock An {Introduction to the Algebraic Approach to some Problems of
  Theoretical P}hysics.
\newblock In: {\em{s{\'{e}}rie des cours et conf{\'{e}}rences sur la physique
  des hautes {\'{e}}nergies}}, No.\,3, 1974.
\newblock Centre de {Recherches Nucl{\'{e}}aires}, Strasbourg.

\bibitem{Uhlm:76}
A.~Uhlmann.
\newblock The transition probability in the state space of a $^*$-algebra.
\newblock {\em Rep. Math. Phys.}, 9:273--279, 1976.

\bibitem{Uhlm:85}
A.~Uhlmann.
\newblock The transition probability for states of $^*$-algebras.
\newblock {\em Annalen der Physik}, 42:524--531, 1985.

\bibitem{Uhlm:95}
A.~Uhlmann.
\newblock Geometric phases and related structures.
\newblock {\em Rep. Math. Phys.}, 36(2/3):461--481, 1995.

\bibitem{Uhlm:98}
A.~Uhlmann.
\newblock Entropy {and Optimal Decomposition of States Relative to a Maximal
  Commutative S}ubalgebra.
\newblock {\em Open Sys. \& Information Dyn.}, 5:209--227, 1998.

\bibitem{Yngv:73}
J.~Yngvason.
\newblock On the {Algebra of Test Functions for Field Operators (Decomposition
  of linear Functionals into Positive Ones)}.
\newblock {\em Comm. Math. Phys.}, 34:315--333, 1973.

\end{thebibliography}

\end{article}
\end{document}